\documentclass[graybox, envcountchap]{svmult}

\usepackage{mathptmx}        
\usepackage{amsmath}
\usepackage{amssymb}
\usepackage{color}
\usepackage{helvet}          
\usepackage{courier}         
\usepackage{dirtree}

\usepackage{makeidx}        
\usepackage{graphicx}        
\usepackage{subfig}

\usepackage{multicol}        
\usepackage[bottom]{footmisc}

\usepackage{hyperref}        
\hypersetup{colorlinks=true,urlcolor=blue,linkcolor=blue,citecolor=blue}

\usepackage[misc]{ifsym}

\usepackage[numbers]{natbib}

\usepackage{xspace}

\usepackage[export]{adjustbox}

\newcommand{\km}{\rm\thinspace km}

\newcommand{\cm}{\rm\thinspace cm}
\newcommand{\yr}{\rm\thinspace yr}

\newcommand{\s}{\rm\thinspace s}

\newcommand{\Msun}{\hbox{$\rm\thinspace M_{\odot}$}}

\newcommand{\Msunpyr}{\hbox{$\Msun\yr^{-1}\,$}}

\newcommand{\keV}{\rm\thinspace keV}

\newcommand{\kmps}{\hbox{$\km\s^{-1}\,$}}

\newcommand{\Zsun}{\hbox{$\thinspace \mathrm{Z}_{\odot}$}}

\newcommand{\psqcm}{\hbox{$\cm^{-2}\,$}}

\makeindex             

\begin{document}


\title{Clusters of galaxies}
\author{Jeremy S. Sanders}
\institute{Jeremy Sanders (\Letter) \at Max Planck Institute for Extraterrestrial Physics, Gießenbachstraße 1, 85748 Garching, Germany \email{jsanders@mpe.mpg.de}
}
%
%

\maketitle

\abstract{
High-spectral resolution observations of clusters of galaxies are a powerful tool to understand the physical processes taking place in these massive objects.
Their hot multi-million-degree X-ray emitting cluster atmospheres, containing most of the baryons in these systems, are enriched to around 1/3 of the solar metallicity.
Therefore, cluster spectra host a variety of spectral lines, in particular, the Fe-L complex around 1~keV typically emitted from cooler systems and Fe-K at 6.7~keV seen from more massive clusters.
The line ratios and continuum are sensitive probes of the temperature distribution of the gas, while the strength of lines compared to the continuum measures metallicity.
With sufficient spectral resolution, the velocity structure can be obtained from line widths and shifts in position.
Using detailed modelling, we can then understand better processes such as feedback from active galactic nuclei, mergers and enrichment, that are taking place in clusters.
}


\section{Introduction}

Many galaxies in the universe are found to be gravitationally bound into objects known as groups and clusters of galaxies.
The richest clusters of galaxies consist of thousands of individual galaxies, with total masses of $\sim 10^{15} \Msun$.
Groups of galaxies are lower mass objects containing fewer galaxies ($\lesssim 50$), although the boundary between clusters and groups is not exact.
In the hierarchical theory of the formation of structure, clusters are expected to lie at the densest regions of the cosmic web, built up by the merger of smaller structures over time (e.g. \cite{Allen11}).

It was noted that a number of the X-ray sources in the sky discovered by the \emph{Uhuru} X-ray observatory were associated with groups and clusters of galaxies \cite{Cavaliere71}.
The discovery of an emission feature around 7 keV in an X-ray spectrum of the Perseus cluster observed using the \emph{Ariel 5} X-ray telescope \cite{Mitchell76} provided key evidence that the X-ray emission from clusters of galaxies was thermal in nature, originating from Fe K-shell transitions within the hot plasma within the cluster. The presence of this line in Perseus and other clusters ruled out non-thermal emission mechanisms as the dominant X-ray emission processes within clusters.

This observed X-ray emission originates from the intracluster medium (ICM), the hot plasma cluster atmosphere, making up the majority of the baryonic matter in these objects.
Despite this, due to their large volume, the density of ICM is relatively low.
The electron density, for example, typically peaks at the centre with values between $10^{-1}$ and $10^{-3}$~cm$^{-3}$.
However, the majority of mass in clusters consists of dark matter, which is only indirectly visible.

Galaxy groups range in temperature from a few $10^6$~K (noting that the typical unit for cluster and group temperatures in X-ray astronomy is keV, where $1\keV$ is approximately $11.6$~MK) to $2-3\keV$.
Clusters range above these temperatures to $\sim 10$~keV.
At the highest temperatures, the majority of the X-ray emission from these hydrogen-rich atmospheres is thermal bremsstrahlung (free-free) emission.
At lower temperatures (below $\sim 2\keV$), particularly in group scale objects, line emission becomes more important.
It should be noted, however, that clusters and groups are not purely isothermal, and temperature substructure can be present.
For example, as we discuss further in Section \ref{sect:cool_cores}, these objects often contain a central cool core where the temperature drops significantly below the cluster ambient temperature.
Many baryonic processes in clusters (e.g. mergers, stripping, feedback) can also affect the temperature distribution.

If the ICM entirely consisted of hydrogen and helium, then there would be little point in doing high-resolution X-ray spectroscopy in order to understand cluster physics.
In fact, the metallicity of the intracluster medium is enriched to around $0.2-0.3 \:\Zsun$ (Section \ref{sect:enrich}), meaning that the ratio of the density of metal ions compared to hydrogen, is roughly 0.2--0.3 times the value found in the sun.
It is the emission lines from the metals in the ICM which enable X-ray spectroscopy and allows the detailed study of a cluster through its hot baryons.

High-resolution X-ray spectroscopy can be used in a number of different ways to study the physics of galaxy clusters, including

\begin{enumerate}
  \item Examining the metallicity of the ICM.
    The majority of metals in the ICM originate from stellar processes.
    Studying them allows us to look at the history of star formation in cluster galaxies and constrain models of metal production, and examine the processes which transport metals  through the ICM.
  \item Measuring the velocity of the ICM.
    Given sufficient spectral resolution, the energy and width of emission lines can be used to measure the velocity distribution of the hot atmosphere.
    Indirect spectral probes, such as resonant scattering, can also be used to constrain the velocity in clusters.
    Velocities are powerful observational probes of physical processes in clusters, such as feedback or mergers.
  \item Studying the temperature distribution of the ICM.
    In addition to the bremsstrah\-lung continuum emission, the relative strength of emission lines is sensitive to the temperature of the plasma.
    Temperature measurements can be used to study many physical processes, including whether there could be cooling of the ICM in the core of a cluster.
  \item Non-thermal physical processes.
    Although the bulk of the cluster emission is thermal in nature, there could be other emission mechanisms taking place, such as charge exchange.
    In addition, more exotic particles in the cluster could give produce emission lines or modulate the spectra of sources passing through the cluster.
\end{enumerate}

\section{Cluster emission processes and spectral models}

\begin{figure}
  \centering
  \includegraphics[width=\columnwidth]{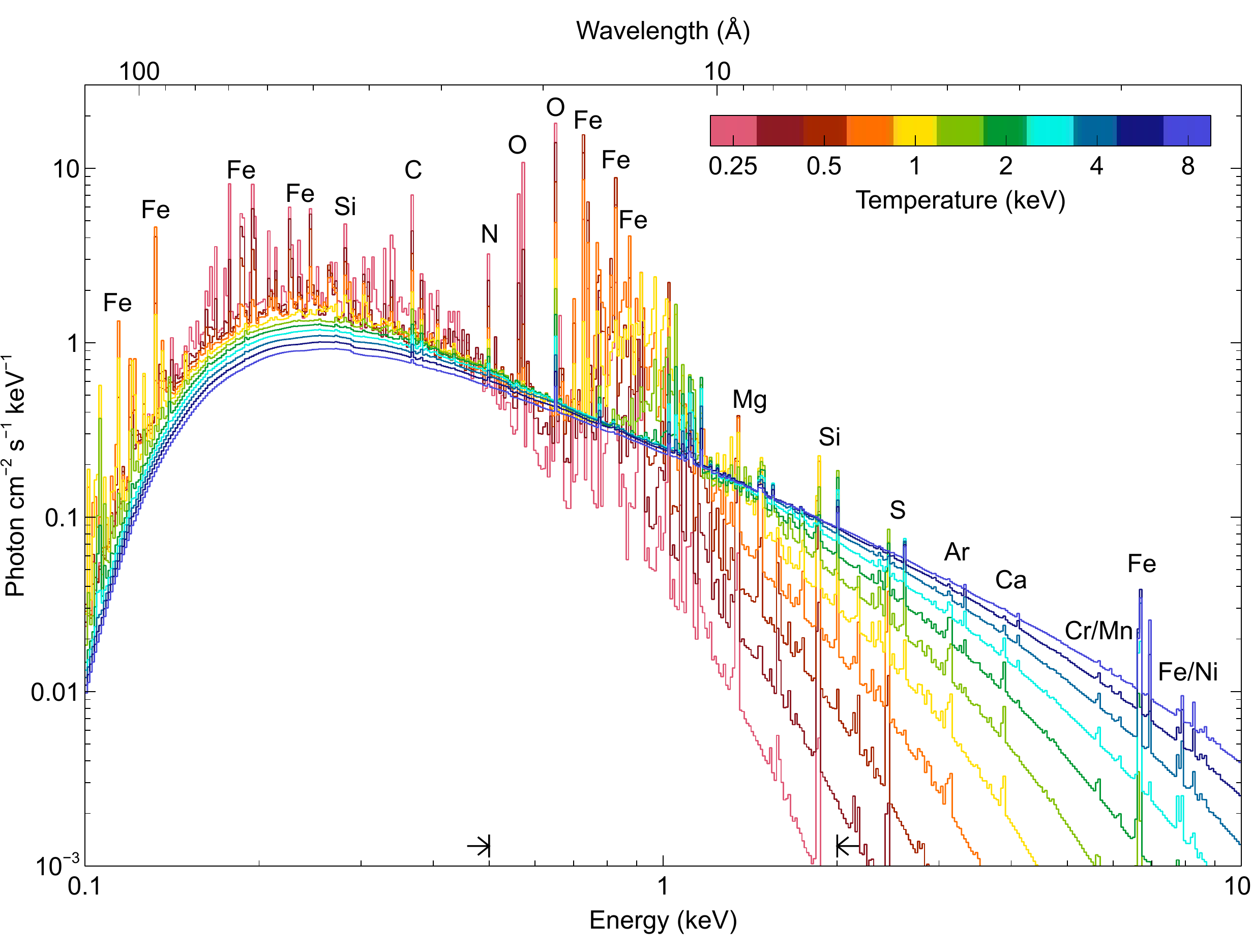}
  \includegraphics[width=\columnwidth]{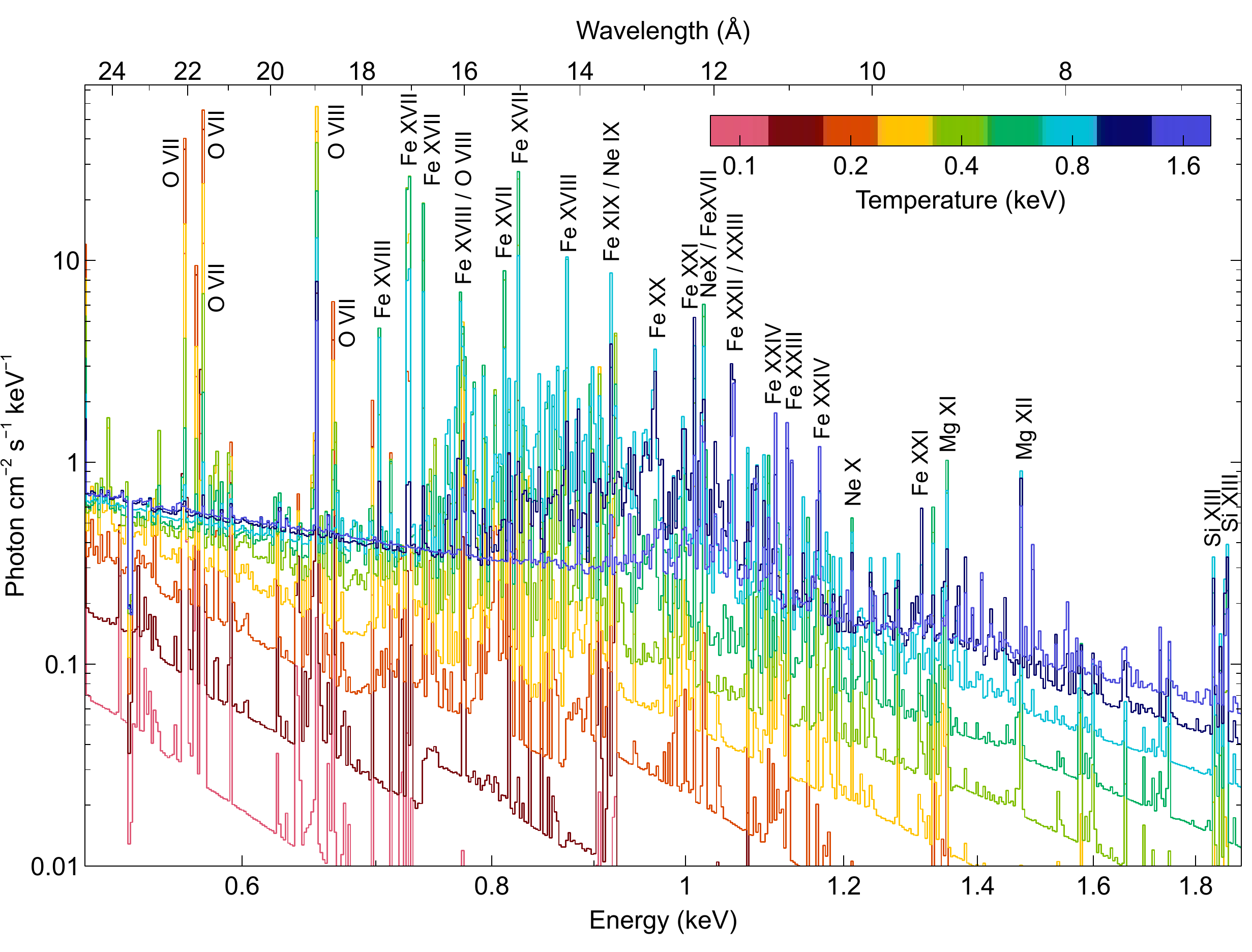}
  \caption{
    Collisionally-ionised spectral models as a function of temperature, for a wider spectral range (top) and around Fe-L (bottom).
    Models were computed with \textsc{apec} 3.0.9 for a $0.3\Zsun$ metallicity, an \textsc{Xspec} normalisation of unity, an equivalent H column absorption of $10^{20} \psqcm$ \cite{Wilms00}, in 500 spectral bins.
  }
  \label{fig:apec}
\end{figure}

The spectra from clusters of galaxies are typically assumed to be well-modelled as a collisionally-ionised optically-thin plasma.
At the temperatures typically found in clusters, the continuum emission is mainly due to bremsstrahlung (free-free) emission (see Chapter 7 for a detailed discussion of astrophysical plasmas, and \cite{Kaastra08} for a review).
In this case the emission is produced by scattering between electrons and ions, with $\epsilon_{ff}$, the rate of emitted photons per unit volume and energy, scaling as
\begin{eqnarray}
  \epsilon_{ff} \propto n_e n_i \: T^{-1/2} E^{-1} e^{-E/kT} Z_\mathrm{eff},
\end{eqnarray}
where $n_e$ and $n_i$ are the electron and ion number densities, respectively, $T$ is the temperature, $E$ is the photon energy and $Z_\mathrm{eff}$ is an effective charge of that ion species.
Bremsstrahlung emission is dominated by hydrogen and helium ions.
Other, smaller, continuum contributions come from free-bound and two photon emission processes.
The cluster X-ray emission is heavily weighted towards the densest regions in a cluster.

For line emission, an ion is put into an excited state, usually by collisional excitation for cluster plasma, which then decays into a lower state by emitting a photon.
This two-body process also scales as the particle density squared, similarly to the bremsstrahlung emission.
However, as the lines are from heavier elements than helium, the number density of an ion depends on the abundance of the relevant element, and the fraction of those atoms which have the correct ionisation, which varies by temperature.
Lines in the X-ray spectrum are strongest for those elements which are most abundant, but the strength also typically increases strongly according to the nuclear charge \cite{Kaastra08}, which results in iron (Fe) lines being some of the strongest in the spectrum.

The assumption of collisional ionisation equilibrium (CIE) means that for a particular temperature, there is a balance between the rate at which ions and atoms are ionised to a higher ionisation state and those recombining from the higher states.
In a non-equilibrium ionisation (NEI) plasma this balance in rates is not present.

The most common codes used for modelling the spectra of a collisionally-ionised plasma in equilibrium are the \textsc{apec} model produced from \textsc{atomdb} \cite{SmithApec01}, and the \textsc{spexact} code in the \textsc{spex} modelling software \cite{KaastraSPEX96} which derives from the earlier \textsc{mekal} model \cite{MeweMekal85,MeweMekal86,LiedahlMekal95}.
New observational data and model results are continually being incorporated into these models.
For example, high resolution \emph{Hitomi} spectra of the Perseus cluster were used to bring these models into better agreement \cite{HitomiAtom18}.
Fig.~\ref{fig:apec} shows some example spectra at different temperatures over the energy range typical for clusters (0.1-10 keV) and in detail the range around the Fe-L complex (0.5-1.9 keV).
The results are for an \textsc{Xspec} normalisation of $1\cm^{-5}$, which is defined for a source with redshift $z$ and angular diameter distance of $D_{A}$ (cm) as
\begin{eqnarray}
  N = \frac{10^{-14}}{4\pi [D_{A} \, (1+z)]^2} \int n_\mathrm{e} n_\mathrm{H} \, \mathrm{d}V,
\end{eqnarray}
integrating the electron ($n_\mathrm{e}$ [cm$^{-3}$]) and Hydrogen ($n_\mathrm{H}$ [cm$^{-3}$]) densities over volume ($\mathrm{d}V$ [cm$^{3}$]).

Fe is the strongest line-producing element in the spectra of clusters of galaxies.
At lower temperatures (below $\sim2$ keV) the Fe L-shell lines (lines from $n=2$ in Li-like to Ne-like iron ions) around 1 keV are dominant.
At higher temperatures the He-like Fe K-shell lines at $\sim 6.7$~keV become the main spectral feature.
Other lines which are important contributors to the X-ray spectrum include those from C, N and O (in the soft sub-keV band), Ne and Mg (in the Fe-L complex range), Si and S ($\sim 2-3$ keV), Ar and Ca ($\sim 3-4$~keV) and Ni (in the Fe-L region and $\sim 7-8$~keV).

Cluster spectra, however, are not perfectly described as perfect CIE optically thin models.
Firstly, an observer has to account for absorption from material in our own Galaxy (e.g. using \textsc{tbabs}; \cite{Wilms00}).
The amount of absorption can be estimated from H\textsc{i} and dust extinction maps of the galaxy \cite{Willingale13}.
High resolution spectra can be used to model the absorption directly, although high-quality data are required (e.g. \cite{Pinto13}).
Emission from our own galaxy and the local hot bubble can also produce a significant background contribution \cite{Snowden97,Liu17LHB}.

The assumption that the plasma is optically thin may also be incorrect for some resonance lines, leading to a modification of the spectral line intensity distribution over the source.
We discuss this effect further, including how it can be used to infer gas motions, in Section \ref{sect:reson}.

There are other components which may need modelling.
The X-ray background and AGN in the cluster can introduce complexities in modelling the cluster spectrum, particularly when determining the level of the continuum.
Charge exchange is another process which can give rise to emission lines and is predicted to happen at the interface between the ICM and colder gas (Section \ref{sect:beyondcie}).
Clusters are also not completely static objects, and so there are cases where there may be a lack of ionisation equilibrium (NEI).
This lack of thermal equilibrium is most likely to occur in shocks in the low-density outer regions of clusters (also Section \ref{sect:beyondcie}).

One of the largest uncertainties in spectral modelling is deciding whether more than one or how many temperature components are sufficient for describing the observed spectrum.
Real clusters have temperature substructure, either due to projection effects or real structures within the studied aperture.
An observer can fit either a single component, two or more components, or a continuous parametrized distribution (a differential emission measure or DEM).
Given a perfect thermal model and data, it should be possible to test a number of distributions and choose the one which best fits the data, but existing data has limited statistics or spectral resolution.
In addition, these components could have different metallicities; it is common for observers to assume they have the same metallicity in order to constrain the parameters.
Using too few components when modelling can, in particular, lead to the determination of incorrect metallicities (the so-called Fe-bias or inverse Fe-bias, see Section \ref{sect:enrich}).
If a DEM is used to model the spectrum then commonly used distributions are powerlaws or a log-Gaussian (e.g. \cite{dePlaa06}).

\section{High spectral resolution data}
In X-ray astronomy, one can either achieve high spectral resolution using a detector with intrinsic high spectral resolution, such as a microcalorimeter or through the use of a diffraction grating.

Extended objects, such as clusters, present some challenges when using gratings.
The main issue is that gratings on X-ray telescopes are slitless, which means that the spectral resolution is degraded for an extended object when compared to a point source.
The relative origin angle of an X-ray photon and its wavelength is both important for the dispersion angle.
For a reflection grating instrument like the Reflection Grating Spectrometer (RGS) gratings onboard the \emph{XMM-Newton} observatory, the wavelength ($\lambda$) is connected to the incoming angle ($\alpha$), outgoing angle ($\beta$), for a spectral order ($m$, which is negative for this type of grating) and grating spacing ($d$) by
\begin{eqnarray}
  m \lambda = d(\cos \beta - \cos \alpha).
\end{eqnarray}
Therefore an angular form of the profile of the source along the dispersion direction is convolved with the X-ray spectrum.
However, a grating does allow the spectral variation to be examined in the cross-dispersion direction, allowing a limited study of the spatial variation of the source spectrum.

For the RGS spectrometers onboard \emph{XMM-Newton}, the spectra are broadened by approximately
\begin{eqnarray}
  \Delta \lambda \approx \frac{0.139}{m} \frac{\Delta \theta}{\mathrm{arcmin}} \: \textrm{\AA},
\label{eqn:broad}
\end{eqnarray}
where $\Delta\theta$ is the half energy width of the source and $m$ is the (positive) spectral order (\emph{XMM-Newton Users Handbook}).

In order to analyse the grating spectrum a measured profile of the object can be included within the spectral model (e.g. using the \textsc{lpro} spectral model in the \textsc{spex} package \cite{KaastraSPEX96,Pinto15}), or a fitting approach can be used to fit for the line broadening caused by the spatial extent (e.g. \cite{SandersRGS08}).
One should note, however, that the source profile may be strongly energy dependent.
For example, if a cluster has a cool core the spectrum of its centre can have strong emission lines but would be continuum-dominated at a larger radius.
An incorrect profile could over or underestimate the spatial broadening, particularly if measured using a low-spectral resolution instrument \cite{SandersVel13}.
A measured spatial profile also includes the instrumental point spread function (PSF) which will need to be accounted for when modelling the spectral broadening.

To obtain the highest quality grating spectrum of a cluster requires an object where the line-emitting region is as compact as possible.
Therefore our grating view of clusters of galaxies is biased towards those with steeply peaked surface brightness profiles.
At higher redshifts, spatial line broadening becomes less of an issue, but the effective area of current grating instruments limits our ability to observe more distant clusters with reasonable exposure times.

\begin{figure}
  \centering
  \includegraphics[width=0.62\columnwidth]{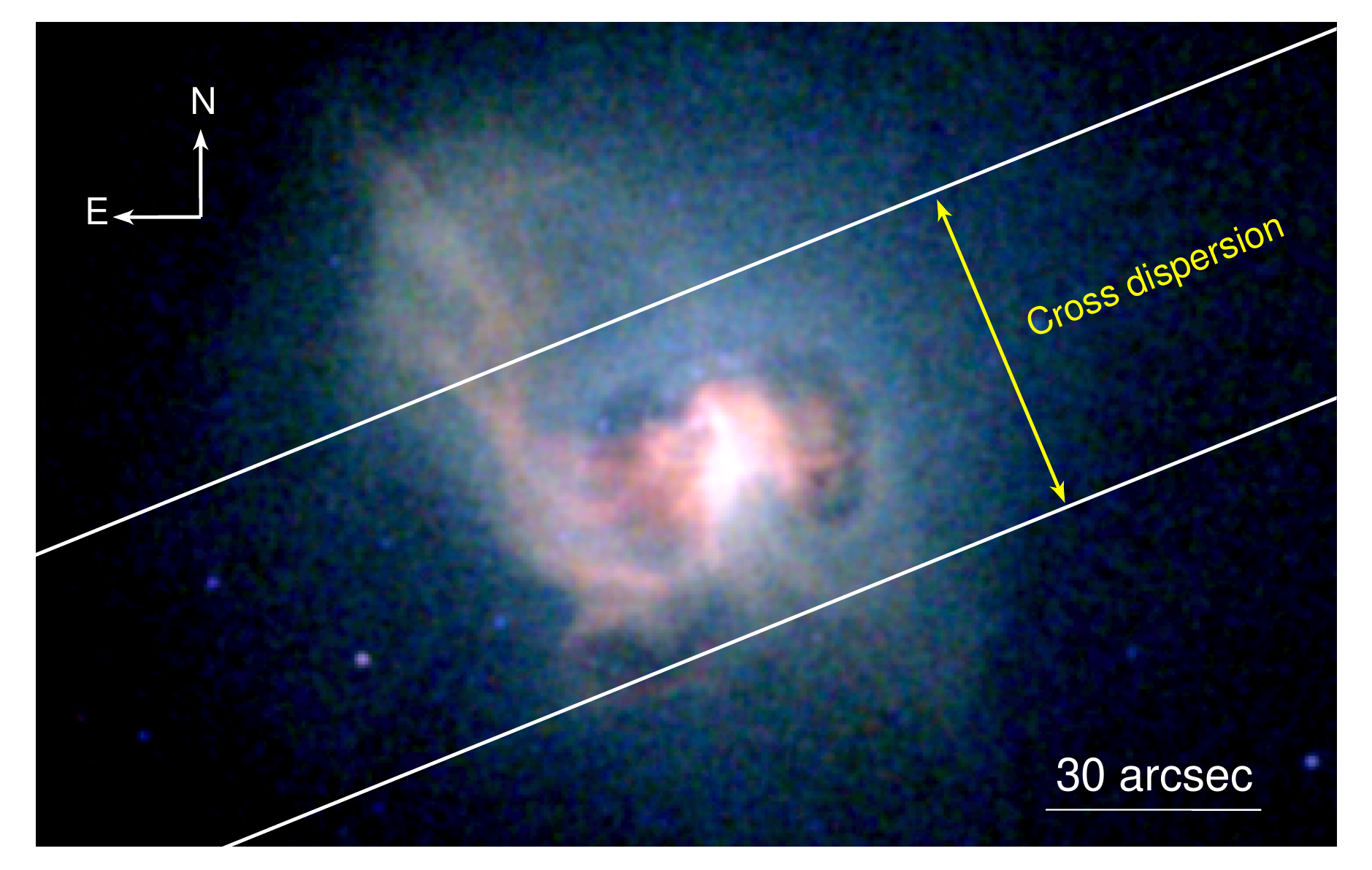}
  \includegraphics[width=\columnwidth]{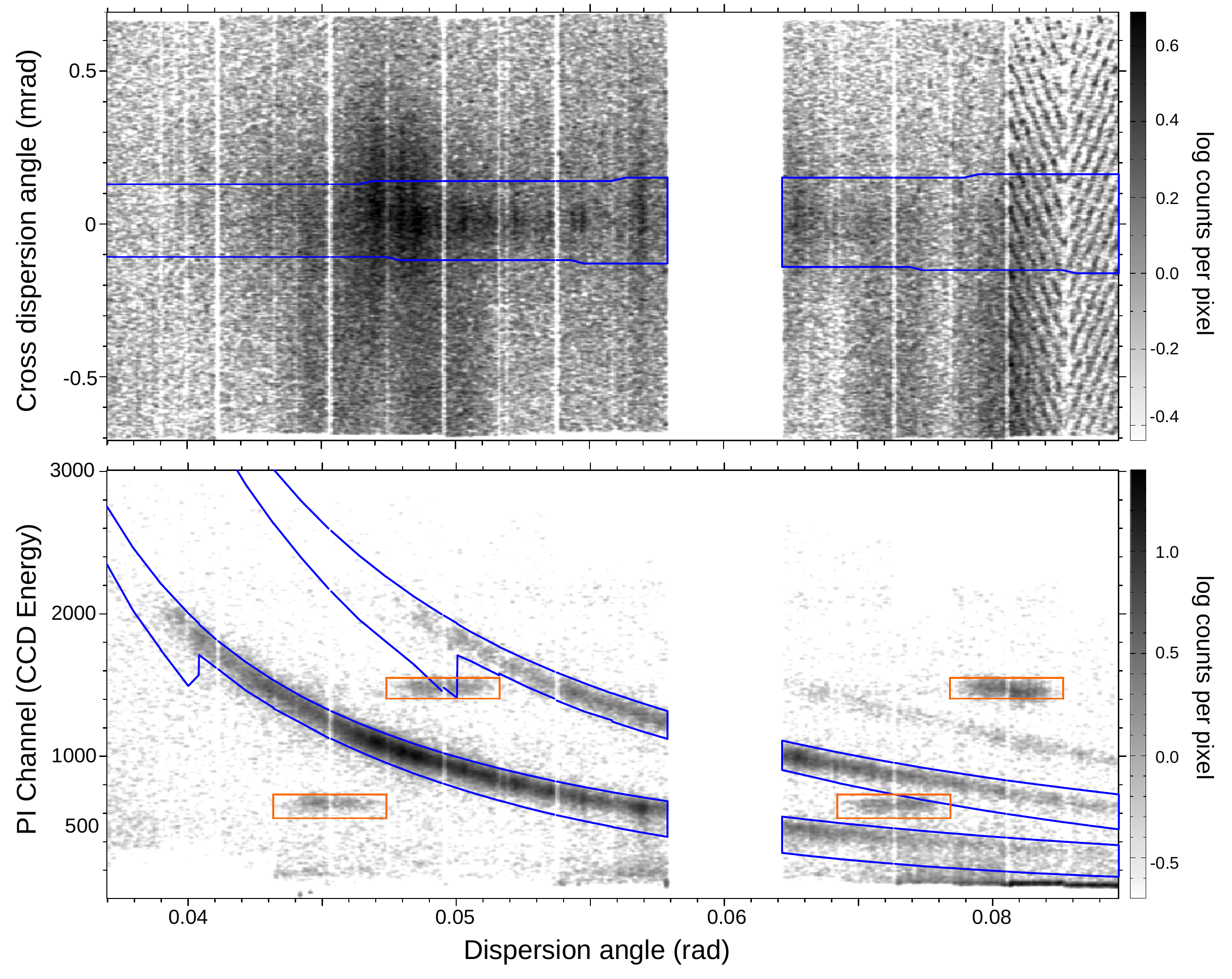}
  \caption{
    \emph{XMM-Newton} RGS observation of the Centaurus cluster of galaxies, taken with the RGS2 camera.
    (Top panel) \emph{Chandra} RGB (0.5-1.0, 1.0-1.5 and 1.5-6.0 keV) image of Centaurus (taken from \cite{SandersCent16}) showing the RGS extraction region.
    (Centre panel) Distribution of the X-ray counts in cross-dispersion and dispersion-angle space, showing the 90\% PSF width cross dispersion extraction region, which corresponds to around 0.8 arcmin in extraction width.
    (Bottom panel) Distribution of counts in the PI channel (pulse-invariant energy as measured by the CCD) as a function of dispersion angle, showing the 95\% pulse-height extraction regions for the first two spectral orders.
    The orange boxes show the calibration sources.
    The \emph{XMM} dataset is observation ID 0406200101 \cite{SandersRGS08}.
    The empty dispersion range is due to a failed CCD in the RGS2 camera.
  }
  \label{fig:banana}
\end{figure}

Fig.~\ref{fig:banana} shows the data obtained from an \emph{XMM-Newton} RGS grating camera for an example source, the nearby Centaurus cluster of galaxies ($z=0.010$, \cite{SandersRGS08}), whose central region contains a complex structure of multiphase gas.
The top panel is a high spatial resolution \emph{Chandra} X-ray image of the source and the approximate RGS extraction region, showing the cross-dispersion axis direction.
The centre panel plots the one-dimensional cross-dispersion image of the source as a function of the dispersion angle.
The bottom panel shows the spectral orders as separated using the energy resolution of the CCD detectors.
To analyse the event file from the camera, the user needs a source position and parameters for the cross-dispersion extraction (parametrized as a percentage of the PSF width) and CCD energy extraction widths.
The extraction width optimises how much of the source to extract and is used to maximize the signal-to-noise ratio.
Increasing the width in the cross-dispersion direction or CCD energy dimensions gives the user more photons, but includes more background.
However, for extended objects, such as clusters, the cross-dispersion width also modifies the part of the source which is being examined in that spatial dimension.
If a cluster has a cool core, like the example shown, a spectrum extracted from a smaller cross-dispersion extraction width will likely contain the emission lines with less background emission.

The dispersion and cross-dispersion directions depend on how the gratings are mounted in the telescope and the roll angle of the telescope on the sky.
Spacecraft operational constraints also decide the rotation of the instrument at the time of the observation, meaning that the angle of the grating on the sky cannot be freely chosen.

Modelling the background components in spectra also presents a challenge for nearby clusters.
For \emph{XMM-Newton} there exists a tool which produces a background spectrum for non-X-ray-backgrounds, based on empty field observations selected using the off-axis count rate.
In addition there will be cluster background components originating from radial regions beyond the radius of the cross-dispersion extraction region, in projection along the line of sight and along the dispersion axis.

A different approach to solving the source extent problem is to use a forward-modelling technique to predict the measured photons in the dispersion-cross dispersion plane and to self-consistently model the source \cite{Peterson01,Peterson07}.
Combining observations with multiple roll angles, or also forward-modelling CCD data could help build up much more accurate 2D models of the source.

Microcalorimeters are much better suited to extended objects as they are intrinsically pixel-based instruments.
If there is more than one pixel in the detector, then different spatial regions can be extracted by selecting which pixels to example.
However, if the PSF of the telescope is relatively broad compared to the pixel size (as in \emph{Hitomi} \cite{HitomiTel16} or \emph{XRISM} \cite{XRISM20}), the pixels will not correspond to independent regions in the cluster.

\section{Cool cores}
\label{sect:cool_cores}
A substantial fraction of the clusters in the local universe have cores which are strong steeply-peaked emitters of X-ray radiation (e.g. \cite{McDonald13}).
This is due to the rising gas density towards the centres of these objects.
As the X-ray emission scales with the square of the density, the X-ray emission is very peaked.
Therefore if one calculates the mean radiative cooling timescale due to the radiation of X-rays, it can be very short, and much shorter than the age of the cluster.
These clusters with dense cooler material in the centre are known as cool core clusters.
Other non-cool core clusters have much flatter density profiles and isothermal cores \cite{Hudson10}, although there is a debate about whether there are two separate populations or a continuous distribution (e.g. \cite{Ghirardini22}).

If there is not some form of heating, then this material should be rapidly cooling out of the X-ray waveband in cool core clusters.
This cold material could lead to a build-up of cool gas in the cores of clusters, and subsequently the formation of stars.
The rate of material cooling out of the X-ray band is called the mass deposition rate ($\dot{M}$), within this cooling flow model \cite{Fabian94}.
$\dot{M}$ can be estimated from the excess X-ray luminosity ($L$) emitted in some regions where the radiative cooling timescale is shorter than the age of the object using
\begin{eqnarray}
  L = \frac{5}{2} \frac{\dot{M}}{\mu m} kT,
\end{eqnarray}
taking account of the gravitational work done on the gas as it flows towards the core,
where $T$ is the temperature and $\mu m$ is the mean mass per particle.
Typical values ranging from 10s to 100s of $\Msunpyr$, although there are extreme cases above $1000\Msunpyr$\cite{McDonald12}.
Despite these expected rates of cooling, the measured star formation rates are a small fraction of the total mass deposition rate (e.g. \cite{Johnstone87,Nulsen87,McDonald18}).

\begin{figure}
  \centering
  \includegraphics[width=\columnwidth]{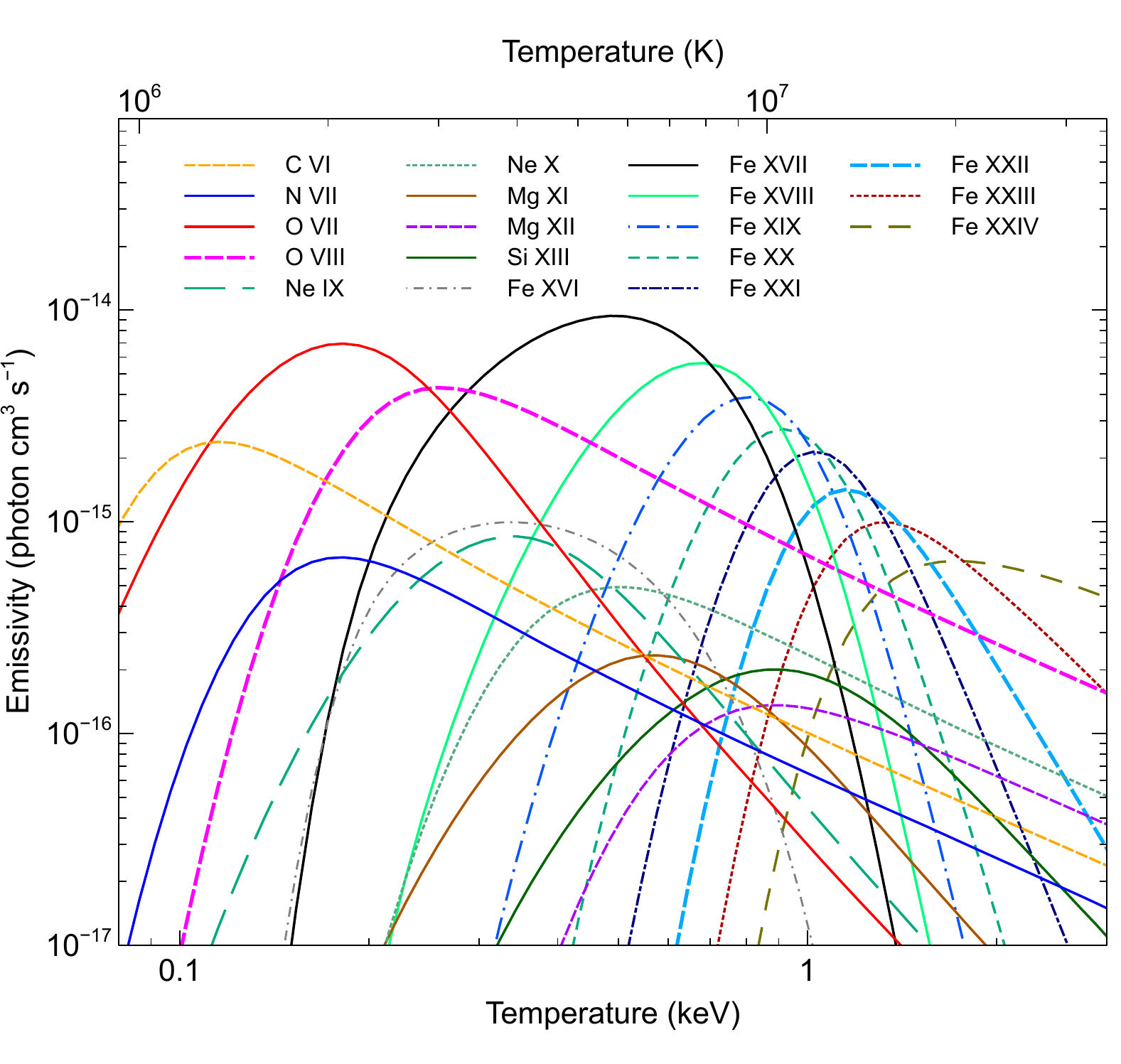}
  \caption{
    Total line emissivities of ions in the RGS waveband (5--38{\AA}) as a function of temperature.
    Emissivity here is defined to be the rate of radiative transitions divided by the product of the electron and H number densities.
    Therefore the relative strength of emission lines is that of a plasma with a constant density as a function of temperature.
    Values were calculated using \textsc{apec} 3.0.9 \cite{SmithApec01} assuming solar abundances \cite{AndersGrevesse89}.
  }
  \label{fig:emissivities}
\end{figure}

In a cooling flow, there should be a range of gas temperatures from some ambient cluster temperature down to temperature values which should not emit X-ray emission.
This is typically modelled spectrally through a multi-temperature component model, where the component normalisations follow a DEM distribution.
The commonly-used \textsc{mkcflow} and \textsc{vmcflow} models in the \textsc{xspec} spectral fitting package \cite{ArnaudXspec} follow the prescription of \cite{Mushotzky88}, where there is some number of temperature components between an upper and lower temperature.
The differential emission measure at a temperature is then set to be inversely proportional to the bolometric luminosity at that temperature.
The combined model spectrum is normalised to measure the mass deposition rate.
If the lower temperature of the model is set to zero, then a full cooling flow is modelled.
One limitation of this model, however, is that it does not explicitly include the gravitational work done on the gas in the cooling flow.

When high resolution \emph{XMM}-RGS grating observations were first made of the core of nearby clusters, the spectra did not show the expected emission lines seen from lower temperature X-ray emitting gas, as predicted from models like the one described above.
The power of these high-resolution spectra comes from a large number of temperature-sensitive emission lines present in the soft X-ray RGS band.
These include Fe L-shell emission lines and others from O, Ne, Mg and Si.
In Fig. \ref{fig:emissivities} is plotted the strength of these emission lines as a function of temperature for Solar abundances.
It can be seen that the presence or absence of some emission lines, or their relative strengths, are very good indicators of gas temperature.
The best indicators at low X-ray emitting temperatures are the Fe~XVII lines, showing gas around 0.5~keV and O~VII, at around 0.2~keV.

In Abell 1835 \cite{Peterson01} it was found that there was a lack of Fe XXIII, Fe XII and Ne X emission, implying there was little material below 2.7 keV in temperature, with less than $200\Msunpyr$ cooling to very little temperatures, compared to the $1000\Msunpyr$ expected from the X-ray profile.
In Sérsic 159-03 \cite{Kaastra01} a two-temperature component fit was sufficient, implying less than 20\% of the predicted $230\Msunpyr$ cooling.
An isothermal model was found to fit the RGS spectrum of Abell 1975 \cite{Tamura01a}.
The lack of Fe XVII emission lines constrained the mass deposition rate to less than $150 \Msunpyr$, compared to the $500 \Msunpyr$ inferred from the X-ray emission.
Abell 496 \cite{Tamura01b} also showed a deficit of material at lower temperatures.

\begin{figure}
  \includegraphics[width=\columnwidth]{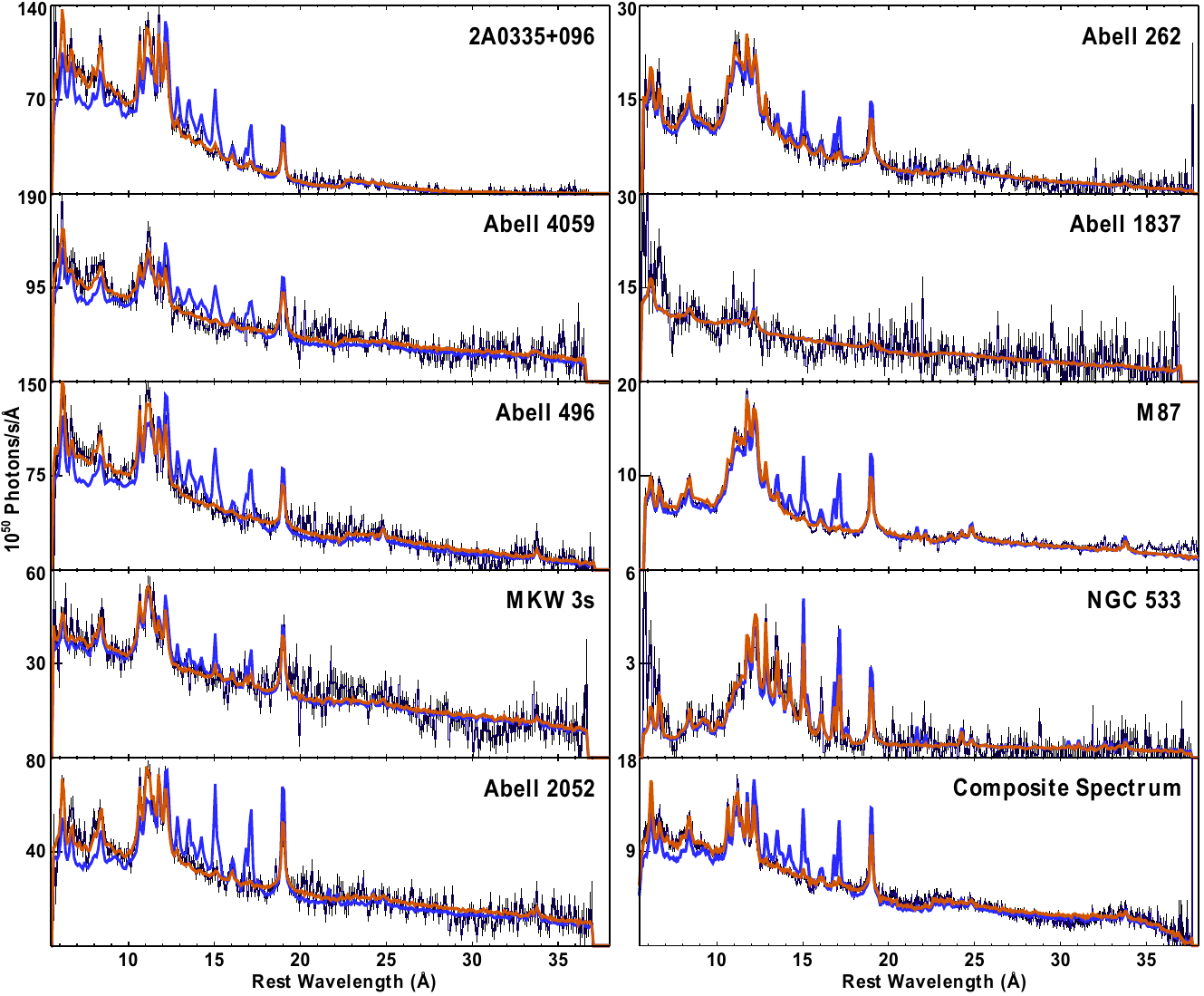}
  \caption{
    Fluxed and de-redshifted RGS spectra of several galaxy clusters and a composite spectrum, taken from \cite{Peterson03}.
    Shown on the data points is the best empirical model (red) and a cooling flow model scaled to match the soft X-ray flux (blue).
    It can be seen that the cooling flow model strongly overpredicts the flux of emission lines from colder gas.
  }
  \label{fig:peterson}
\end{figure}

Later studies examined more targets or looked at deep observations in detail.
In an analysis looking at a sample of multiple clusters \cite{Peterson03}, 14 targets with high mass deposition rates were studied, finding that the emission lines indicated gas typically down to around 1/2 of the ambient temperature, but very little below 1/3 of that temperature.
Fig.~\ref{fig:peterson} shows the spectra for a subset of these clusters, compared to cooling flow models with the appropriate mass deposition rates.

There is good evidence that this lack of cooling is due to the AGN feedback in these systems (for a review see \cite{Fabian12}).
In many of these clusters the central AGN are seen to be actively affecting their surroundings by the mechanical input of energy (as an example, see the Centaurus cluster in Fig.~\ref{fig:banana} top panel).
The radio jets of these AGN form bubbles of radio-emitting plasma in the intracluster medium.
The mechanical heating power can be estimated using both an estimate of the enthalpy of the bubbles (using the surrounding gas to measure the pressure) and a timescale for the formation of the bubbles \cite{Churazov01}.
There is a good correlation between the rate of X-ray cooling in the centre of clusters and this heating power (e.g. \cite{Rafferty06,HlavacekLarrondo12}), implying that the AGN are responsible for preventing a large fraction of the cooling.
The details of how the energy is dissipated from the feedback into the ICM are, however, uncertain.
Studying the cool X-ray emitting gas in cluster cores and the associated multiphase material constrains models of AGN feedback.
There are scenarios such as the chaotic-accretion model \cite{Gaspari13Chaotic}, where thermal instabilities cause cold clouds to `rain' onto the black hole, or another where cold material can condense behind buoyantly rising AGN-inflated bubbles \cite{McNamara16}.

\begin{figure}
  \centering
  \includegraphics[width=\columnwidth]{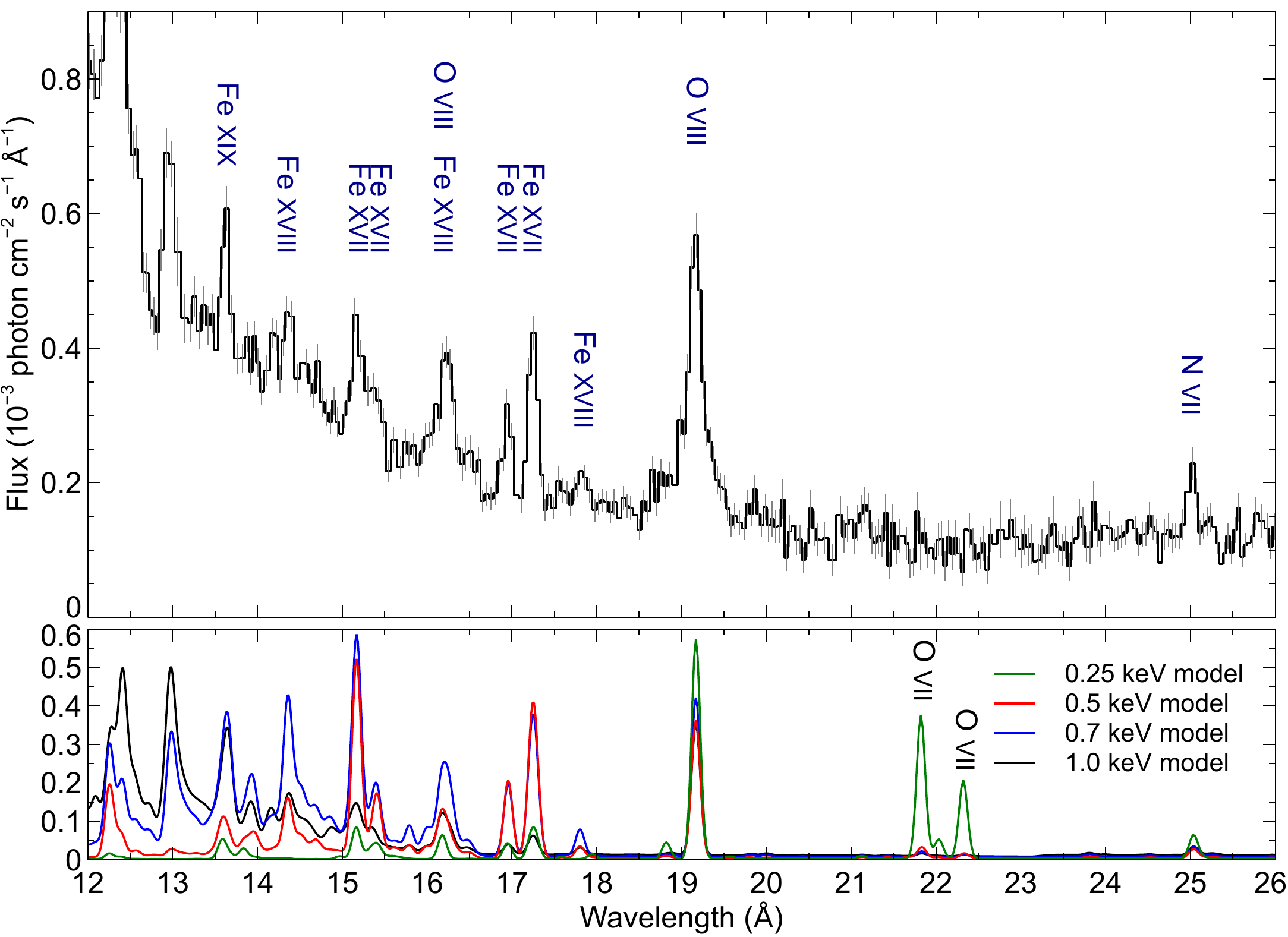}
  \caption{
    Fluxed RGS spectrum of the core of the Centaurus cluster of galaxies, extracted from the innermost 0.8 arcmin width, taken from \cite{SandersRGS08}.
    The bottom panel shows smoothed \textsc{apec} models at various temperatures for comparison.
    The O~VII emission lines, indicating material around 0.25~keV, are weak in the observed spectrum.
  }
  \label{fig:censpec}
\end{figure}

\begin{figure}
  \centering
  \includegraphics[width=0.49\columnwidth]{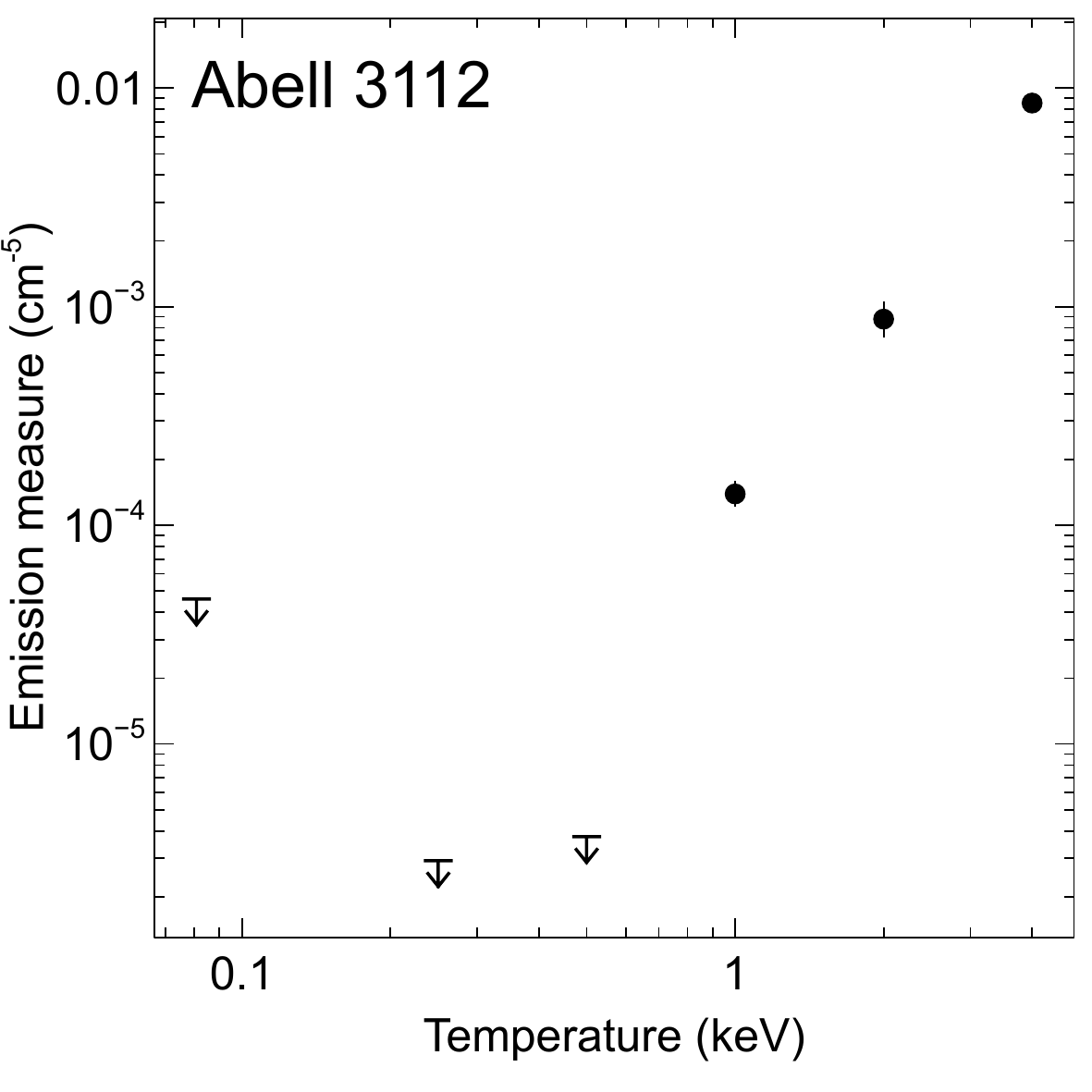}
  \includegraphics[width=0.49\columnwidth]{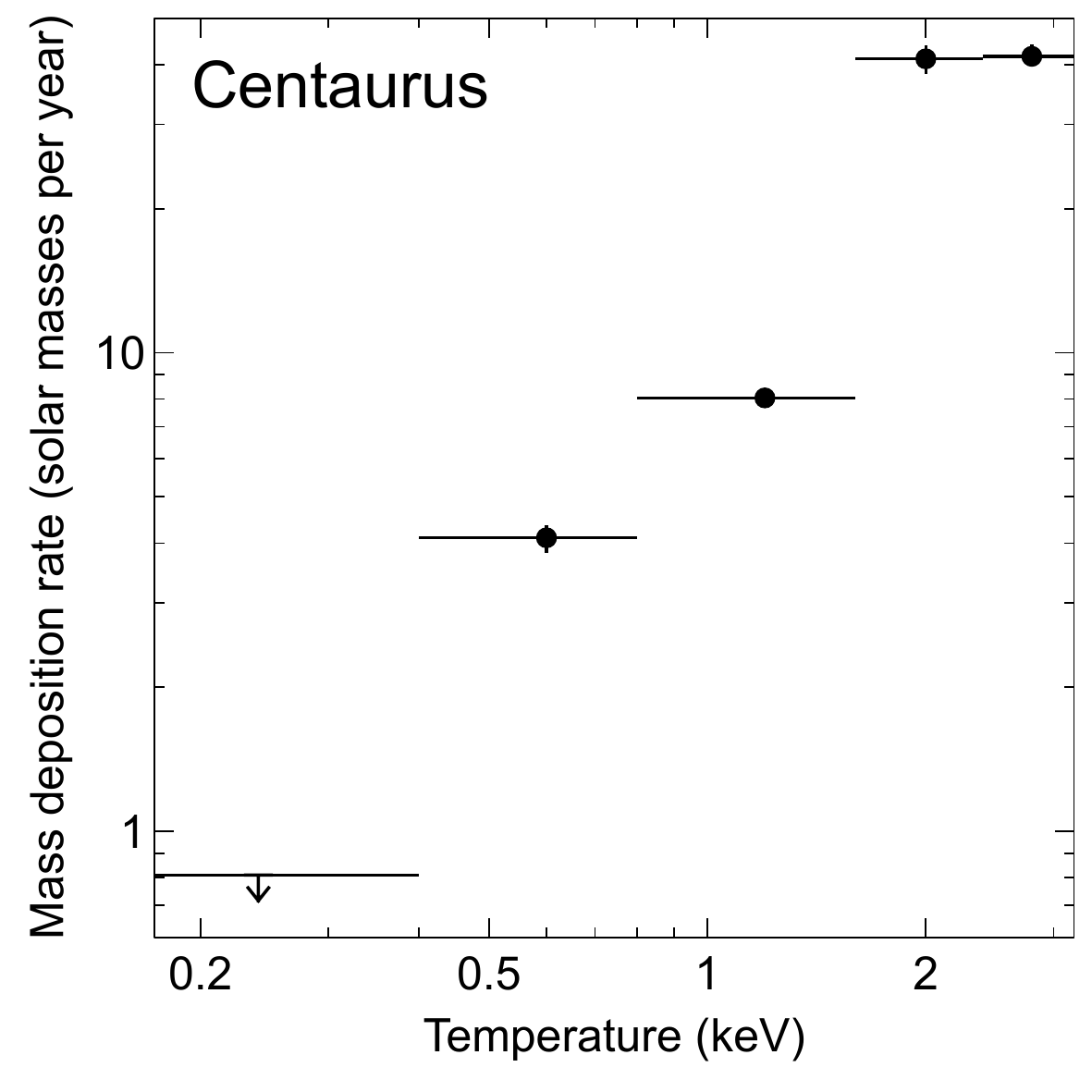}
  \includegraphics[width=0.98\columnwidth]{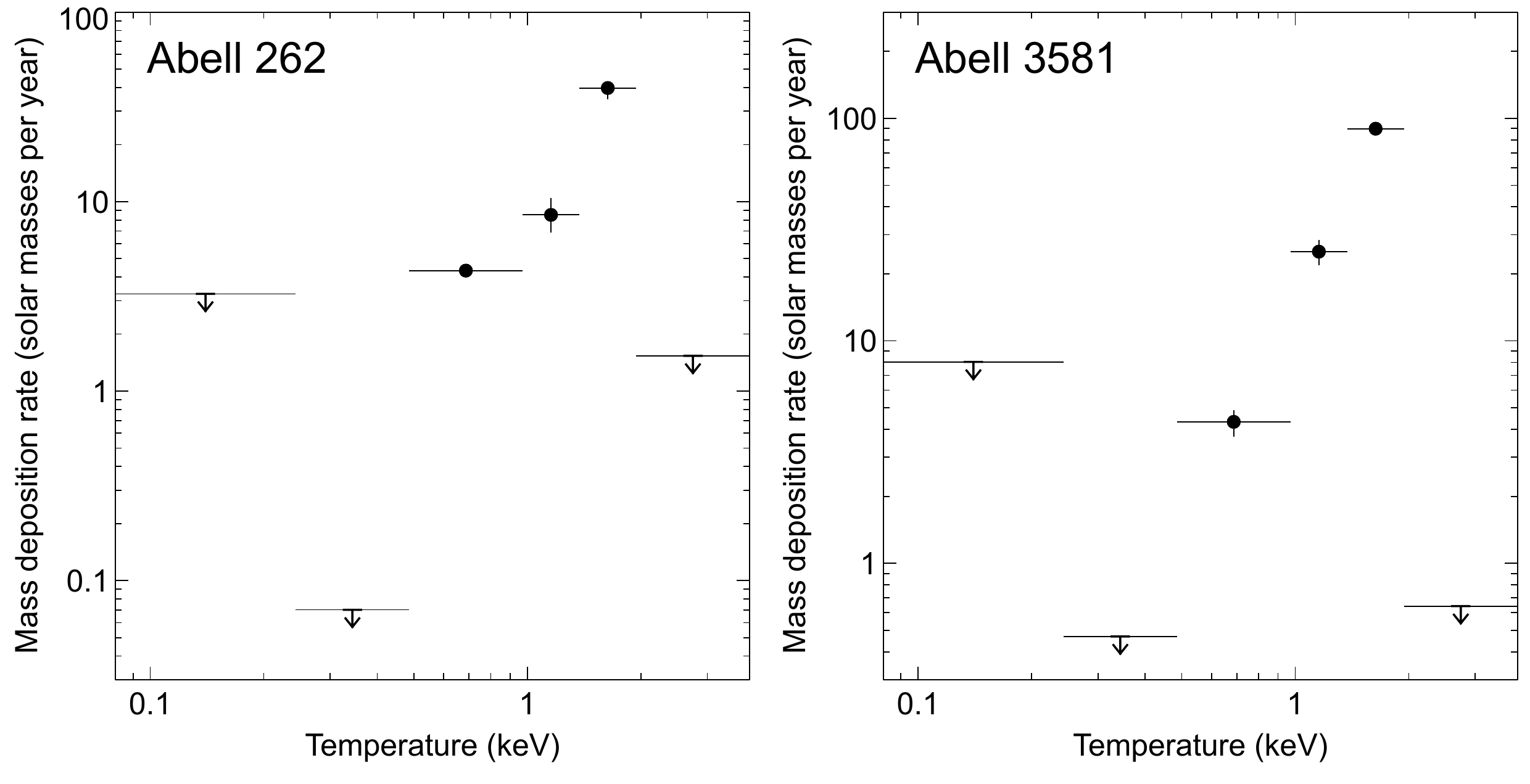}
  \caption{
    Temperature distributions in four clusters, measured using deep \emph{XMM}-RGS observations.
    For Abell 3112, the emission measure for 6 \textsc{apec} components is shown (values taken from \cite{Bulbul12}).
    For Centaurus (values from \cite{SandersRGS08}), Abell 262 and Abell 3581 (values from \cite{SandersRGS10}) the data have been modelled using \textsc{vmcflow} components where the temperature range has been split into 5 or 6 sub-ranges, plotting the mass deposition rate.
  }
  \label{fig:tempdist}
\end{figure}

Deep observations of clusters allow tight constraints to be put on the level of cooling that can take place.
For example the Centaurus cluster \cite{SandersRGS08} shows cooling down from above 3 keV down to around 0.5~keV, as shown by the presence of the Fe~XVII emission lines in the spectrum (Fig.~\ref{fig:censpec}).
However, there is no strong evidence for O~VII emission in the spectrum, although a reanalysis found evidence for weak emission \cite{Pinto14}, measuring the cooling levels below 0.25~keV temperature consistent with the strength of the lines.
The amount of possible mass deposition as a function of temperature can be measured by fitting models consisting of multiple temperature components, approximating a DEM, or multiple cooling flow models spanning the temperature range.
The results for Centaurus and three other clusters are shown in Fig.~\ref{fig:tempdist}.
At temperatures below 0.5 keV there are often sharp cut-offs in the temperature distribution, here constrained by the lack of O~VII emission.
However, in lower mass group-scale objects there is some evidence for gas at still lower temperatures, as seen by O~VII emission in stacked observations \cite{Sanders11_OVII} or in individual systems \cite{Pinto14,Pinto16}.

The very low levels of allowed cooling found in the centres of these systems are suggestive that there is a tight connection between cooling and feedback.
There cannot be strong overcooling or overheating.
This conclusion can be somewhat relaxed if the coolest material can be hidden or cooled more rapidly than expected, for example by mixing.
The very coolest X-ray emitting material could mix with cool gas associated with the H$\alpha$ nebulae common in such systems \cite{Fabian02}, such as in the Virgo cluster \cite{Werner10}.
In a large sample of clusters and groups, truncated cooling flow models were fit to the RGS spectra \cite{Liu19}.
If the model was allowed to cool to zero, then the mass deposition rate was around 10-30\% of the classical $\dot{M}$.
However, if they terminate at 0.7 keV in clusters, the cooling rates are higher.
Energetically the cooling gas could power the emission-line nebulae found in these systems.
There also remains the possibility that absorption intrinsic within the cooling region could obscure substantial amounts of cooling material \cite{Fabian22}.



\section{Motions in galaxy clusters}
\subsection{Introduction}
Measuring motions in the ICM is a very important probe for many different physical phenomena in clusters.
However, such measurements are very difficult to make due to the very high spectral resolution necessary to detect typical motions of $100-1000\kmps$, either in terms of the shifts due to bulk motions or increases in line widths due to velocity distribution.
Although there are some existing results, it will not be until the launch of microcalorimeters such as those onboard \emph{XRISM} and \emph{Athena}, that such measurements will become more common.

One particularly important physical process which generates motions in the ICM is the AGN feedback in the cores of clusters.
The feedback should induce motions of few hundred $\kmps$ in the ICM (e.g. \cite{Heinz10,BourneSijacki17}).
By measuring the velocity distribution around the central AGN we can measure the kinetic energy of the gas, testing feedback models.
The level of energy in turbulence, shocks and sound waves can be compared to the thermal energy of the ICM and other non-thermal components.
This ``complete-calorimetry'' of cluster cores will be very powerful for understanding AGN feedback \cite{CrostonAthena13}.
For these purposes, ideally, is needed a telescope and detector able to resolve the spatial structures around the AGN cavities.

There are other sources for bulk motions and turbulence in a cluster.
Clusters are not static objects within the hierarchical formation of structure.
Merging substructures will induce shocks, flows and turbulence in the ICM.
The influence of this activity is strongest in the outskirts of the cluster \cite{Lau09,Vazza11}.
In addition to generating turbulence, merging subclusters can cause the gas to slosh inside the potential well due to the gravitational centre moving away from the location of ICM peak during the merger \cite{Ascasibar06,ZuHone18}.
These sloshing motions can be several hundred $\kmps$ and last for Gyr.

Measuring the motions in clusters is not only important for our understanding of cluster evolution and AGN feedback.
Non-thermal pressure support affects the assumption of hydrostatic equilibrium often made to measure cluster masses \cite{Nagai07}.
Therefore, it is useful to measure how close to equilibrium clusters are in different dynamical states.
The strength of motions is also an indirect indicator of poorly measured properties of the plasma in the ICM, such as its viscosity and magnetic field substructure \cite{ZuHone15}.
Flows and turbulence in clusters also affect how metals injected into the ICM are propagated through it over time \cite{RebuscoDiff05}.

\subsection{RGS line widths}
Even though the \emph{XMM}-RGS have reduced spectral resolution for extended objects, it was found that they could be used to place constraints on the line of sight velocity width of the ICM.
Although the spectra are broadened by the spatial extent of the source, the upper limit of the line width is still an upper limit to the velocity width.
However, the technique is limited in practice to those clusters with a compact cool line emitting core, producing emission lines in the RGS spectral range and reducing spatial broadening, meaning that it works best on cool-core clusters.
This technique was first demonstrated in Abell~1835, with a 90 per cent upper limit of $274 \kmps$ \cite{Sanders10_A1835}.
It was later applied to other clusters, such as Abell~3112 \cite{Bulbul12}, and samples of clusters \cite{SandersVel11}.
In addition, the spatial extent of the source was also later included in the analysis, either by measurements from imaging or modelling the spectra with different components \cite{SandersVel13,Pinto15}.

\begin{figure}
  \includegraphics[width=\textwidth]{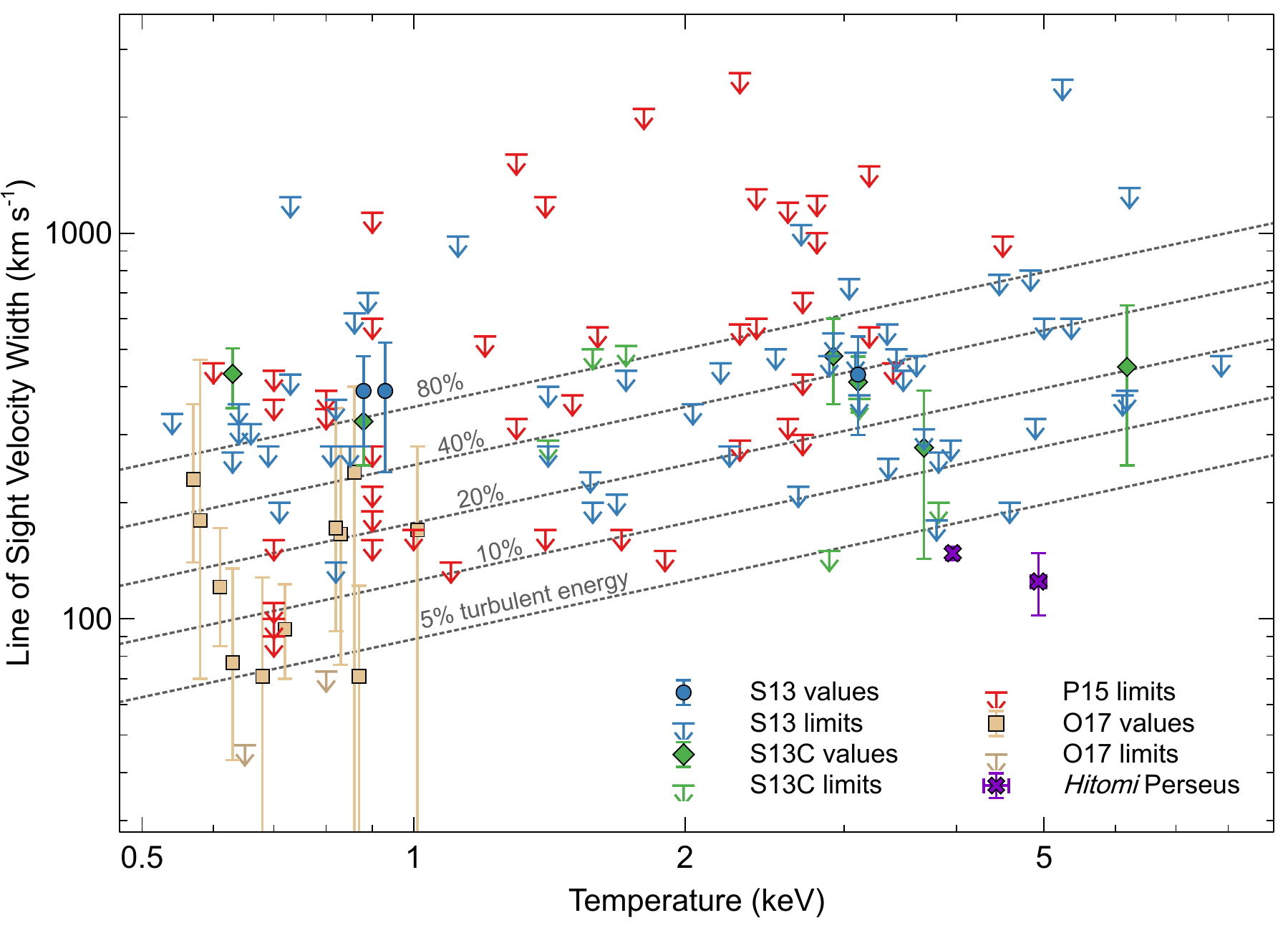}
  \caption{
    Compilation of cluster X-ray velocity width upper limits (using $1\sigma$ confidence levels for homogeneity) and measurements obtained using different samples and techniques.
    The S13 values are for a sample of clusters and groups \cite{SandersVel13}, based on RGS line widths.
    This analysis also includes modelling the spatial extent of the sources, so there are a few points with detections (detections below $2\sigma$ are shown as upper limits).
    The S13C values are for a subset of the same sample, using the spatial profile obtained using narrow-band \emph{Chandra} imaging.
    In this analysis, if more than one value was obtained for some clusters we show the most uncertain of these.
    P15 points are upper limits based on the CHEERS sample of clusters \cite{Pinto15}, taking their conservative values which fit for spatial broadening.
    Upper limits and measurements based on an analysis of elliptical galaxies incorporating modelling of resonant scattering \cite{Ogorzalek17} are shown as O17.
    The \emph{Hitomi} velocity dispersions for the inner and outer pointings of Perseus \cite{HitomiDynamics18} are also plotted, taking the temperatures from \cite{HitomiTemp18}.
    Plotted are lines of a constant fraction of turbulent energy to thermal energy.
    We note that some sources are present in more than one of these sets of results.
  }
  \label{fig:vellimit}
\end{figure}

Fig.~\ref{fig:vellimit} shows a compilation of results from different papers showing mostly upper limits and a few detections of velocity broadening.
The results of S13 and P15 are based purely on the line widths, whereas O17 also applies resonant scattering models to their elliptical galaxies, which we will shortly discuss.
The \emph{Hitomi} result is discussed in Section \ref{sect:hitomi_v}.
Also included in the plot are lines showing levels of turbulent energy if the line of sight width ($v_{1D}$) is assumed to be due to isotropic turbulence, using the relation \cite{Ogorzalek17}
\begin{eqnarray}
  v_{1D} = \sqrt{
    \frac{\varepsilon_\mathrm{turb}}{\varepsilon_\mathrm{therm}}
    \frac{kT}{\mu m},
  }
\end{eqnarray}
where $\varepsilon_\mathrm{turb} / \varepsilon_\mathrm{therm}$ is the fraction of energy in turbulence compared to the thermal energy.
The best upper limits are only $200 \kmps$ and less than $10\%$ of energy in turbulence compared to thermal energy.
The few detections are of the order of a few hundred $\kmps$.

The physical size of the regions probed are different in each cluster and the achievable upper limit is strongly connected to the brightness of the emission lines (the presence of colder gas), the compactness of the emitting region (related to redshift) and the length of the observation.
Due to the mixture of physical extraction region sizes any motions in the cluster will be generated by a mixture of the possible mechanisms generating the velocity, including AGN feedback, sloshing and merger-induced turbulence.
The RGS velocities are comparable to or exceed those of the optical line emitting gas often found in these systems (e.g. \cite{Hatch07,CrawfordFabian92}).
The results suggest relatively low amounts of turbulence in the cores of these systems, despite the presence of AGN feedback in their cores.
It indicates that AGN feedback is likely a gentle continuous process.

\subsection{Resonant scattering}
\label{sect:reson}
Another way to constrain the velocities in clusters indirectly is by looking for resonant scattering \cite{Gilfanov87}.
In the absence of motions, the optical depth at the energy of resonance lines in the X-ray spectra of clusters can become significant.
In this scenario, the scattering will produce a deficit of emission in the scattering region, likely where the gas is densest, and an enhancement at a larger radius.
The optical depth of a line scales inversely by its velocity width \cite{Mathews01}, including both any thermal and turbulent components.
Therefore, the relative strength of a resonance line compared to the expected optically-thin prediction can be used to infer the velocity width of a line and any turbulent motion.
The use of resonance scattering is still challenging, however, as it depends on sufficient optical depth in the cluster (as predicted in the case of no turbulence), an accurate model of the optical depth as a function of radius, a good understanding of the atomic physics of the relevant lines (and those lines the resonance lines are compared to) and high spectral resolution.
Resonance lines important in clusters, groups and elliptical galaxies include the He-like Fe XXV resonance line and the $15.01${\AA} Fe XVII resonance line.

\begin{figure}
  \sidecaption
  \centering
  \includegraphics[width=0.6\columnwidth]{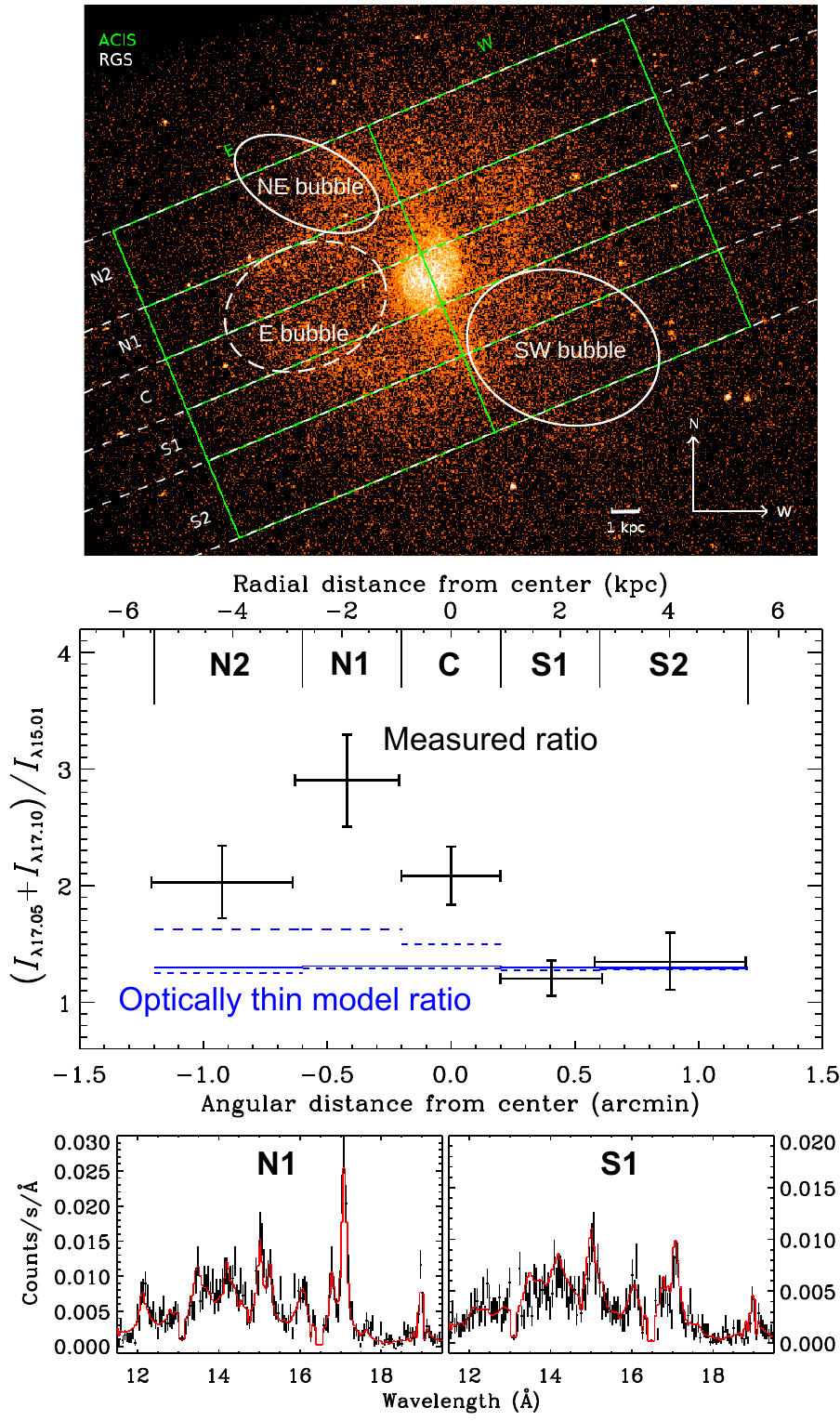}
  \caption{
    Ratios of the strength of non-resonance to resonance Fe XVII line as a function of position in NGC 4636, taken from \cite{Ahoranta16}.
    (Top panel) \emph{Chandra} image of the system highlighting the locations of cavities generated by AGN feedback and marking the RGS extraction regions (from top to bottom, N2, N1, C, S1 and S2).
    (Centre panel) Ratio of the Fe XVII non-resonance lines to the resonance line for each of the extraction regions in black.
    The expected optically-thin ratio is marked as a solid blue line, while the ratios for the coolest (upper) and hottest (lower) temperature components are dashed.
    (Bottom panel) Zoomed-in spectra for two of the regions, showing the different relative strengths of the 15 (resonance) and 17{\AA} (non-resonance) lines.
  }
  \label{fig:ngc4636_res}
\end{figure}

Resonant scattering was found in the giant elliptical galaxy NGC 4636 \cite{Xu02}.
This galaxy was examined in more detail with four other systems \cite{Werner09}, finding low ($<100\kmps$) velocities in these systems and turbulent pressure support $<5$\%.
Two other systems \cite{dePlaa12} were also analysed with line width limits, finding evidence for significantly higher motions ($>320\kmps$) in NGC~5044, which is a merging system.
It is also noted by these authors that there are systematic uncertainties in the strength of the Fe~XVII resonance line compared to the other lines, which limits the use of such measurements (see \cite{HitomiRes18} for a more recent discussion).
Similarly, a sample of 13 systems was analysed, combining resonance scattering and line widths \cite{Ogorzalek17}, finding a best-fitting mean turbulent velocity of $110\kmps$ and a typical non-thermal pressure contribution of $\sim 6$\%.
We plot the results of this analysis on our compilation Fig.~\ref{fig:vellimit}.

Detailed investigations of these systems show the interpretation of measurements of resonance scattering can be challenging.
Fig.~\ref{fig:ngc4636_res} shows RGS spectra from different regions NGC~4636 and an image demonstrating the strong effect of AGN feedback on the gas distribution \cite{Ahoranta16}.
If the ratio of the resonance to a non-resonance line is measured in different cross-dispersion extraction regions, it can be seen that the values vary considerably.
In some regions, it is consistent with the optically-thin case, while other regions indicate resonance scattering.
These differences could be due to observation angle changes towards the AGN jet, or turbulence generated by core sloshing.
Producing a 3D model of the scattering in such an object to make measurements of the gas velocity distribution would be difficult.

\begin{figure}
  \centering
  \includegraphics[width=0.7\textwidth]{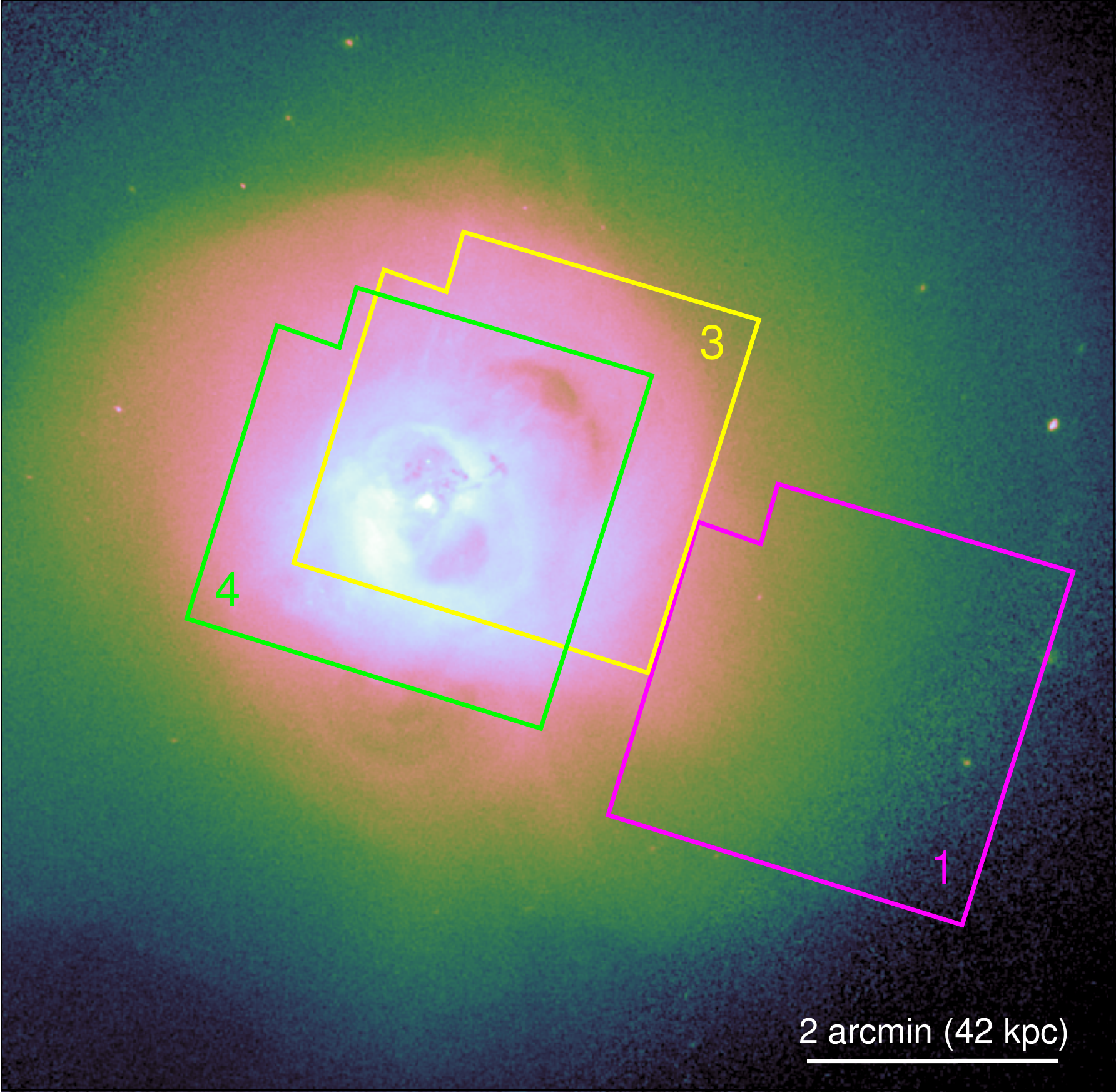}
  \includegraphics[width=0.7\textwidth]{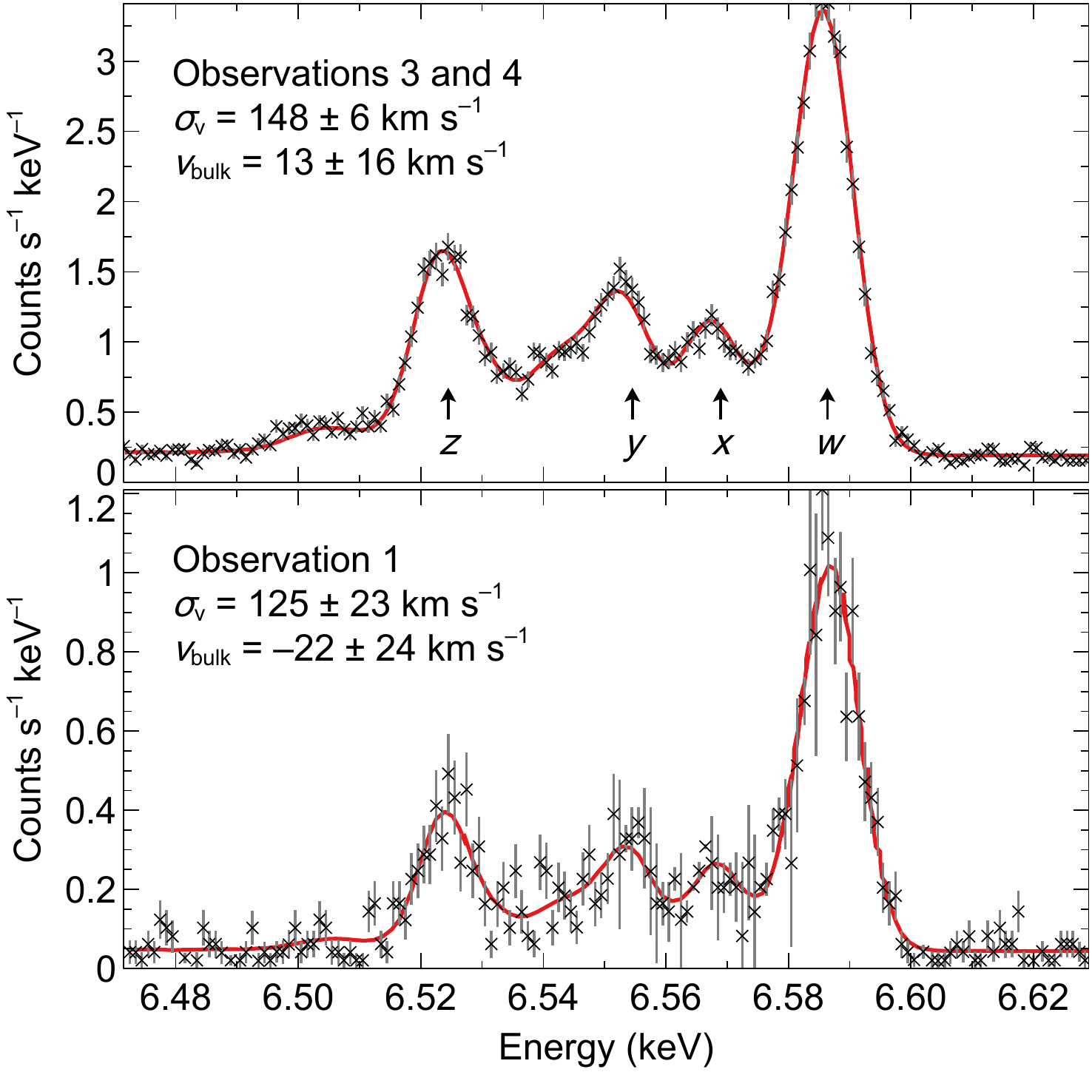}
  \caption{
    \emph{Hitomi} measurements of the velocity line broadening ($\sigma_\mathrm{v}$) and bulk velocity ($v_\mathrm{bulk}$) for different regions of the Perseus cluster (data points from \cite{HitomiDynamics18}).
    The bulk velocity is measured relative to the optical redshift of NGC~1275.
    (Top panel) 0.5-7.0 keV \emph{Chandra} image of the cluster showing the observation footprints.
    (Centre panel) Spectrum around the He-like Fe XXV triplet extracted from two observations (3 and 4) pointed near the innermost region of the cluster, containing the two central cavities and the cavity to the northwest.
    The plot shows the data points and best-fitting model.
    The labelled energies are at those of the resonance (\emph{w}), forbidden (\emph{z}) and intercombination (\emph{x} and \emph{y}) lines.
    (Bottom panel) A spectrum taken from an offset pointing around 80 kpc to the southwest of the cluster centre, close to a cold front edge which may be generated by the sloshing of gas in the cluster potential.
  }
  \label{fig:hitomi_per}
\end{figure}

\subsection{\emph{Hitomi} microcalorimeter results}
\label{sect:hitomi_v}
The launch of the \emph{Hitomi} X-ray observatory \cite{HitomiTel16} with its microcalorimeter detector array promised much improved measurements of the velocity distribution in clusters.
Due to its high ($\sim 5$~eV) spectral resolution in the Fe-K band, it had the capability of measuring line widths and shifts accurately to a high precision.
In addition, being a non-dispersive instrument, the spectra could be obtained from independent regions (although the angular resolution was relatively large at 1.7 arcmin).
Unfortunately, due to the loss of the satellite, only one cluster was observed by \emph{Hitomi}, the Perseus cluster.
These few observations, however, provided a wealth of information.

Fig.~\ref{fig:hitomi_per} shows the spectra extracted from three observations of the cluster (\cite{Hitomi16,HitomiDynamics18}; note observations 3 and 4 are combined)
The data allow the velocity to be measured extremely accurately in these two locations in the cluster.
The pixels in the detector also allow more detailed measurements as a function of position, when the PSF is modelled.
The authors find a velocity gradient of around $100\kmps$ across the cluster core, consistent with sloshing of the gas.
The velocity dispersion is typically around $100\kmps$, with a maximum of around $200\kmps$ towards the AGN and the north-western cavity, likely generated by an earlier episode of AGN activity.
These low-velocity dispersions are similar to what was previously found in the RGS measurements of velocities in clusters.
The line width indicates kinetic pressure support is less than 10\% of the thermal pressure, if isotropic.

\emph{Hitomi} also found evidence for resonance scattering in the resonance line ($w$) in Perseus \cite{HitomiRes18}.
The level of scattering is consistent with the directly measured velocities, although there remain significant statistical and systematic uncertainties, including the uncertainties in model line emissivities.

These low values of velocity width put constraints on our models of how AGN feedback works.
For example, turbulence does not effectively radially propagate if line widths are low \cite{Fabian17,Bambic18,Liu21}.
Some other mechanism is required to radially transport energy in order to prevent the formation of a cooling flow.

\subsection{CCD measurements of bulk flows}
In addition to the high spectral resolution data from \emph{Hitomi} there have been previous measurements of bulk flows in clusters using CCD-type detectors.
If the energy scale of the detector is known sufficiently well, the Fe-K lines can be used to measure velocities.
The accurate energy calibration of \emph{Suzaku} produced hints of motions in samples of clusters \cite{Tamura14,Ota16}.
Recently, \emph{XMM-Newton} pn detector data been examined, making use of instrumental background lines to improve the understanding of the energy scale \cite{SandersPnVel2020}.
The initial results for Perseus agreed with \emph{Hitomi} for a common region and found evidence for sloshing in the cluster.
The gas in the Coma cluster appears to have a bimodal distribution with the velocities of its subclusters.
The technique was similarly applied to the Centaurus and Virgo clusters to look for the effect of feedback on motions \cite{Gatuzz22a,Gatuzz22b}.
CCD instruments have the advantage of large fields of view compared to microcalorimeters.

\section{Enrichment}
\label{sect:enrich}
The intracluster medium is enriched by the stellar processes within the cluster galaxies and the galaxies in the subclusters which merged to form it.
In clusters the ICM is enriched to around 1/3 of the Solar metallicity (for a review see \cite{Mernier18}).
Cool core clusters show a peak towards Solar values in their cores \cite{Mernier17} and a flat distribution at $0.3\Zsun$ in the outskirts \cite{Werner13,Urban17}.
Non cool core clusters do not show this central peak \cite{DeGrandiMolendi01}.
The radial profiles are best measured in the Fe element, but profiles in several other elements (e.g. O, S, Si, Ar, Ca and Ni) show similar shapes \cite{Mernier17}.
In the central part of some cool core clusters there is evidence for a drop in metallicity (e.g. \cite{SandersCent02,Panagoulia13,Mernier17}).

The different stellar processes which make up the abundance pattern in a cluster produce different distributions of elements.
C and N are produced in asymptotic giant branch (AGB) stars \cite{Herwig05}.
Core collapse supernovae (SNcc) produce other lighter metals ranging from O to Mg \cite{Nomoto13}.
In contrast, Type Ia supernovae (SNIa), although their progenitor objects are not confidently identified, nor their explosion mechanism pinpointed exactly, are responsible for heavier elements between Cr to Ni (e.g. \cite{Lach20}).
Intermediate mass elements between Si and Ca could be generated from either SNIa or SNcc supernovae.

The majority of baryonic mass in a cluster is in the form of the ICM and this ICM ends up of a large repository of metals that have been ejected or stripped from the member galaxies.
Therefore studying the metallicity of the ICM is an important tool for understanding these stellar metal-generating processes.
By obtaining the metallicity of different elements in the cluster and comparing the results to models produced by different mixtures of SNIa and SNcc models, these models can be tested.
High-resolution X-ray spectroscopy is the best tool for measuring the ICM abundances.

However, most of the analyses of the radial profiles have been conducted using relatively low spectral resolution CCD instruments, with higher resolution spectra mostly concentrating on the cluster centres.
The measurement of metallicity is made difficult by a number of systematic issues.
One potential problem which is particularly important for lower spectral resolution data is the so-called Fe-bias (for details see \cite{Mernier18}), which occurs in cooler group scale systems, and inverse Fe-bias, which is seen in intermediate temperature systems.
In both of these cases, the problem is that the total observed spectrum originates from more than one temperature component.
If only a single thermal model is fitted to the spectrum, some average temperature is obtained.
In the case of Fe-bias, the average best-fitting temperature causes the best-fitting model metallicity to be reduced to fit the flatter spectral shape around the Fe-L complex.
The inverse Fe-bias similarly increases the best-fitting metallicity to compensate for the inadequate modelling.

\begin{figure}
  \centering
  \includegraphics[width=\columnwidth]{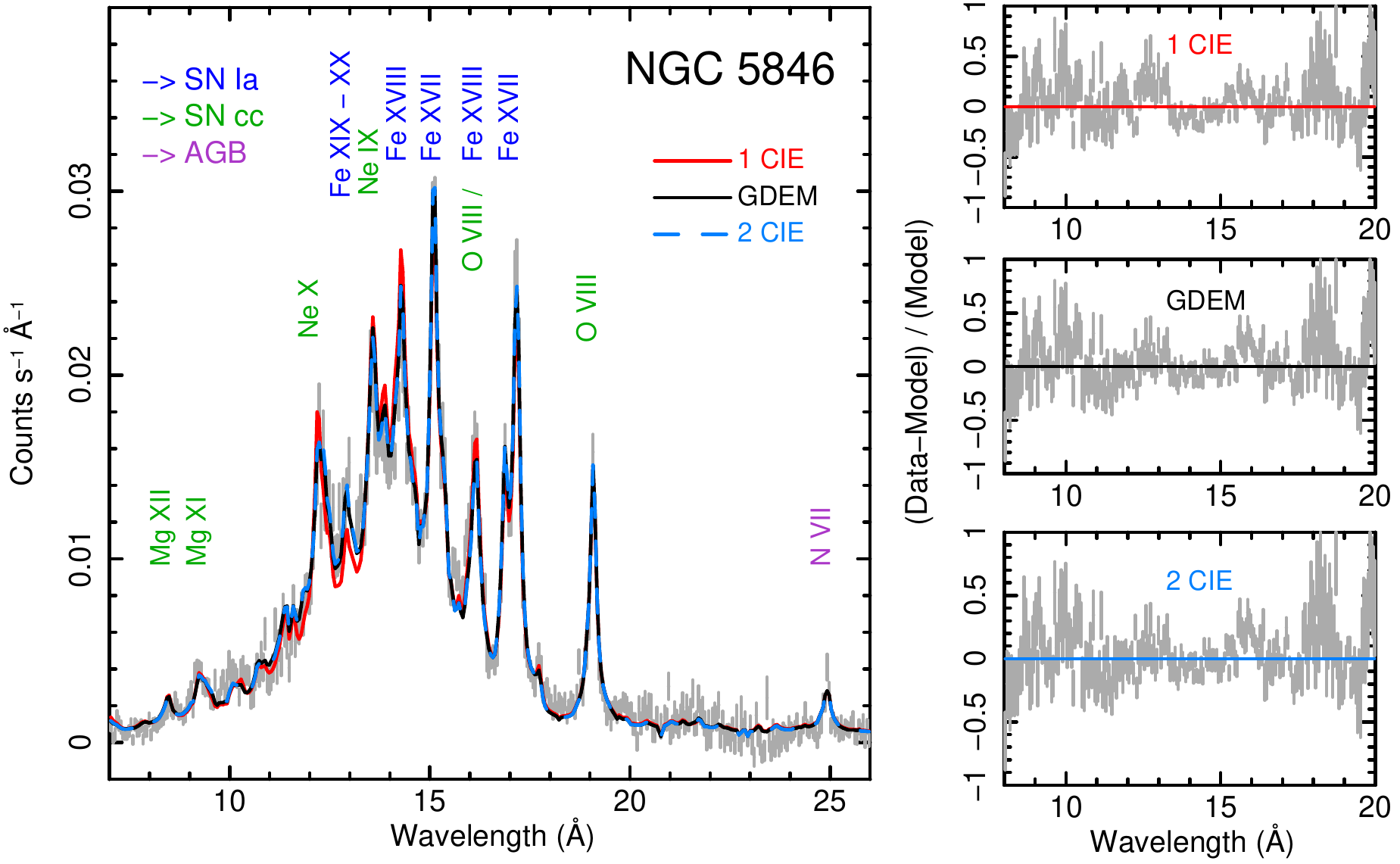}
  \caption{
    RGS spectrum of the NGC 5846 galaxy group, taken from \cite{dePlaa17}.
    (Left panel) Data with models of three different temperature distributions, a single component CIE model, a Gaussian a log Gaussian distribution (GDEM) and a double CIE model.
    The spectral lines are labelled, and coloured according to the most likely origin of their element, i.e. SNIa, SNcc or AGB stars.
    (Right panels) Residuals of the three different models.
  }
  \label{fig:ngc5846_z}
\end{figure}

This problem is much easier to avoid when using higher spectral resolution, but it still remains to some degree in RGS-quality spectra.
As seen in Fig.~\ref{fig:banana} RGS spectra can contain a considerable amount of multitemperature structure due to the large extraction size.
The exact shape of the distribution of temperatures in the extraction region is difficult to predict.
Fig.~\ref{fig:ngc5846_z} shows an example spectrum where the data are fit either by one or two thermal (CIE) components or by a log-Gaussian DEM distribution.
The residual panels show no clear choice of model based on the emission lines.
Simulations predict, for example, biases of 10-20\% on the O/Fe ratio from the uncertainties in the temperature distribution modelling \cite{dePlaa17}.

With RGS spectra the continuum can also be uncertain.
This is important as metallicity measurements need both an accurate line and continuum strength.
Lines from the hotter gas at larger radius in of the cluster can become very broad (due to its large spatial extent), leading to difficulty in modelling it.
In addition, there can be AGN contributions, both from AGN in the cluster and in the X-ray background which should be accounted for in modelling.
Another effect which may bias the level of the continuum is the uncertainties of modelling the absorption towards the cluster, affecting the continuum at lower energies.
A physical effect which could change the continuum is He-sedimentation, which can occur if He were to sediment to the cluster centre over its lifetime \cite{FabianPringle77}.
Although this build-up can be suppressed by various processes, it could affect X-ray metallicity measurements and other cluster properties (e.g. \cite{PengNagai09}).

As we have previously discussed in Section \ref{sect:reson} the spectral models are not perfect, which introduces uncertainties in the line strengths and therefore the obtained metallicities.
Furthermore, resonant scattering could affect abundance measurements, as shown in Perseus where its effect is 11\% for Fe \cite{HitomiRes18}.

\begin{figure}
  \includegraphics[width=\columnwidth]{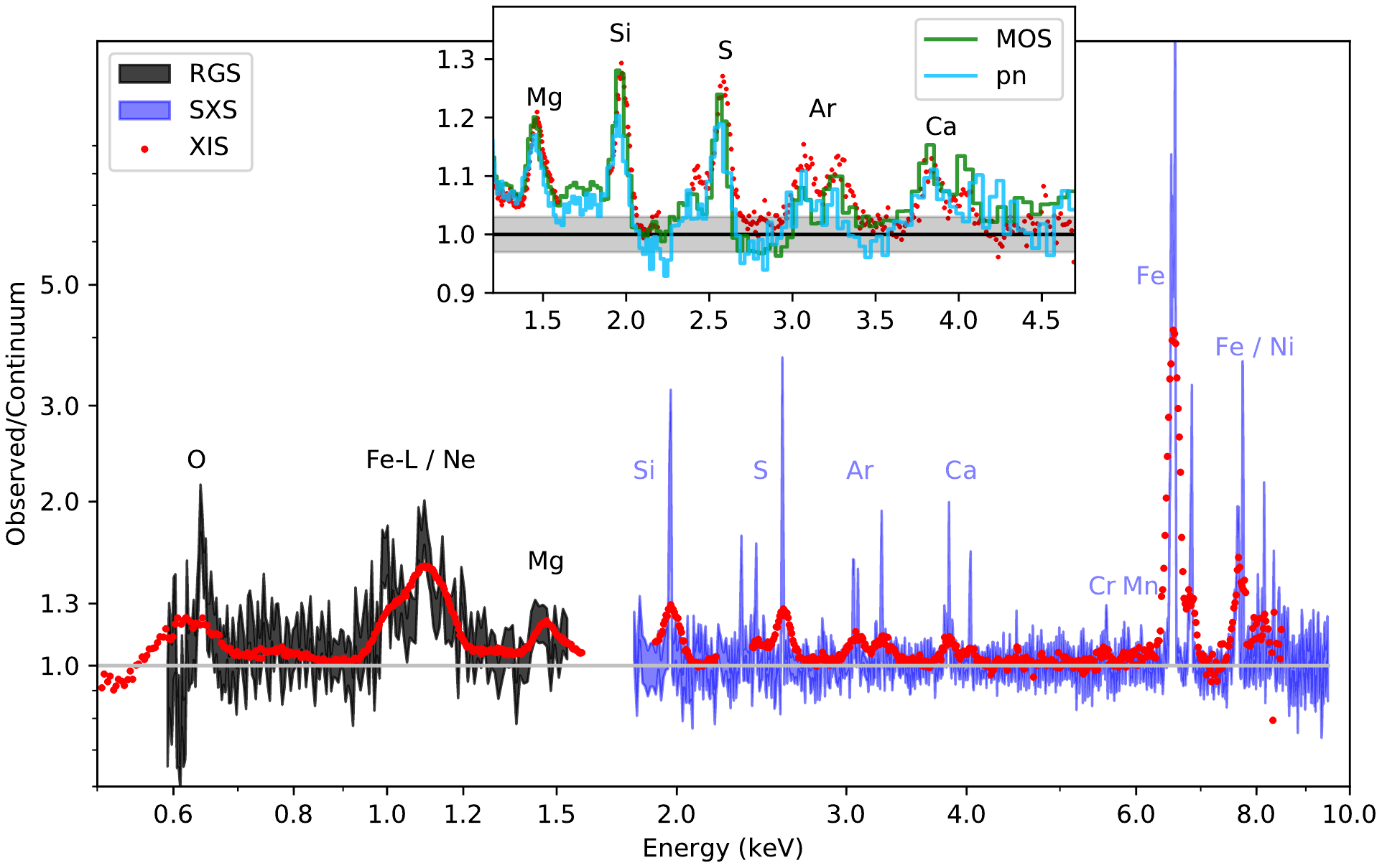}
  \caption{
    Spectra of the core of the Perseus cluster, after having divided by continuum models, taken from \cite{Simionescu18}.
    Results are shown for the \emph{XMM-Newton} RGS, \emph{Hitomi} SXS microcalorimeter and \emph{Suzaku} XIS CCD imaging spectrometer.
  }
  \label{fig:hitomi_z}
\end{figure}

Microcalorimeters, such as the one that was onboard \emph{Hitomi} will make large advances in the measurement of metallicities, due to the lack of spatial broadening, ability to study individual regions and spectral resolution.
Fig.~\ref{fig:hitomi_z} shows a combination of the spectra of Perseus using
the microcalorimeter data at higher energies (soft energies are missing due to a valve in-place during these observations), RGS data in soft energies and lower resolution \emph{Suzaku} XIS data over the whole band.
The extreme improvement of the microcalorimeter over the CCD data is immediately apparent.

\begin{figure}
  \centering
  \includegraphics[width=0.9\columnwidth]{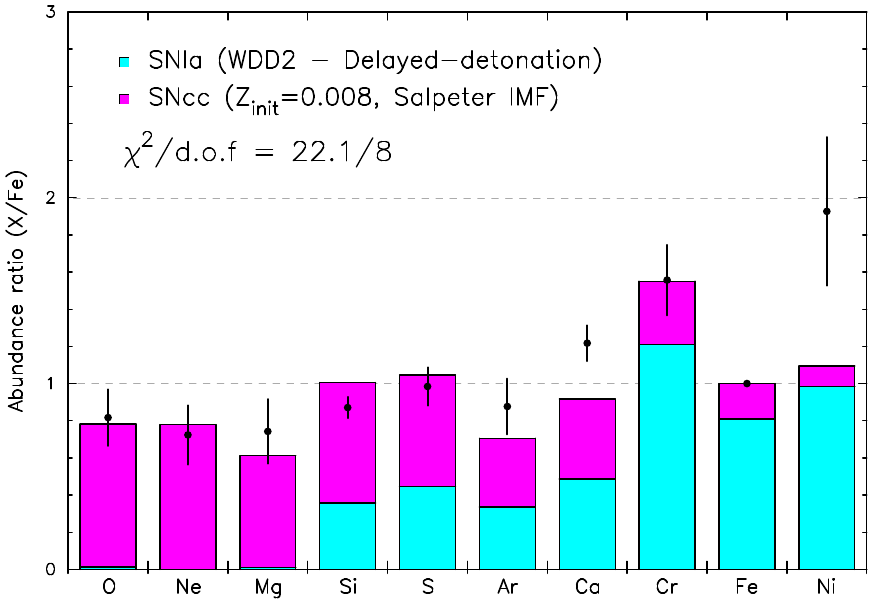}
  \includegraphics[width=0.9\columnwidth]{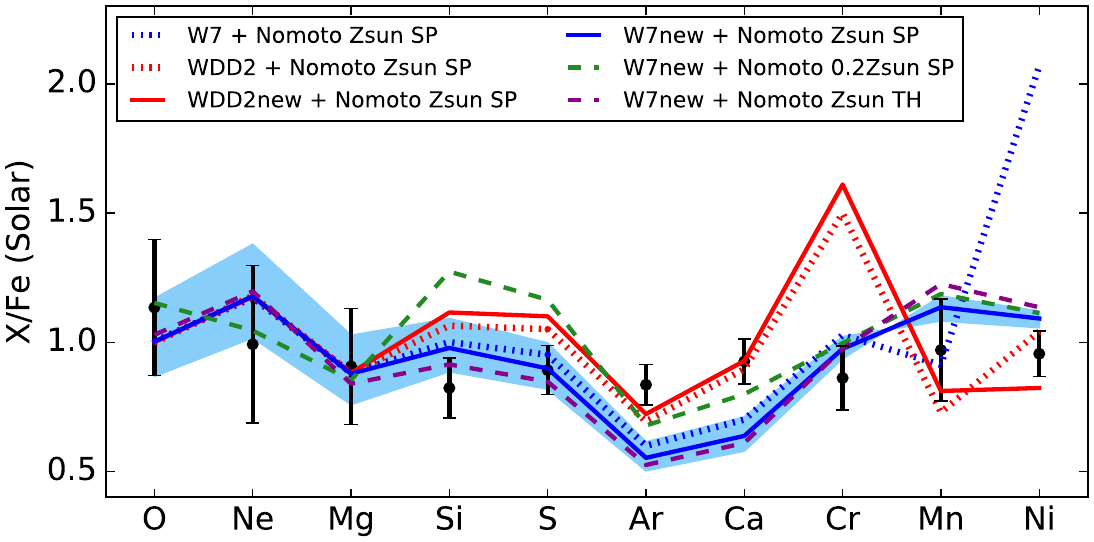}
  \caption{
    Metal abundances compared to Fe.
    (Top panel) Average abundance ratios for the CHEERS sample, taken from \cite{Mernier16b}.
    The O/Fe and Ne/Fe were measured using RGS data, while the others use lower resolution CCD data.
    They  are compared to a best-fitting model of made up of contributions from a classical SNIa model (WDD2) and a SNcc model.
    (Bottom panel)
    The ratios relative to Fe from the \emph{Hitomi} spectrum of the Perseus cluster, or using RGS for the lighter elements (O, Ne, Mg), taken from \cite{Simionescu18}.
    The obtained ratios are compared to various combinations of SNIa and SNcc models.
  }
  \label{fig:snfrac}
\end{figure}

Supernova models can be fitted to the ratios obtained.
Different supernova models can be compared by seeing which matches the data better.
For example, Fig.~\ref{fig:snfrac} shows two sets of ratios and model comparisons.
The top panel is from a sample of 44 clusters, groups and giant elliptical galaxies called CHEERS \cite{dePlaa17}, which were optimised for RGS measurements.
The contributions of SNIa and SNcc to the best-fitting model are shown by coloured bars.
It can be seen that the measured ratios are fairly similar to Solar (X/Fe=1) for many of the elements, except for Ca, Cr and Ni.
Another interesting finding is that stacked RGS spectra \cite{Sanders11_OVII} and individual analyses \cite{Mao19} show enhanced N ratios compared to Solar values.

In Fig.~\ref{fig:snfrac} (bottom panel) is shown the same set of ratios, obtained from high spectral resolution \emph{Hitomi} data from the Perseus cluster, but using RGS for the light elements which could not be measured in this early spectrum \cite{Simionescu18}.
In this case, all of the metals are extremely similar to Solar abundances ratios, with a simple exactly-Solar model giving a $\chi^2=10.7$ for 10 degrees if freedom.
These ratios are remarkably similar to those found in the protosolar nebula, low-mass early-type galaxies and in near-solar metallicity Milky Way stars.
However, this abundance pattern is difficult to reproduce with linear combinations of supernova models, as seen by the mismatch between the data and the models in Fig.~\ref{fig:snfrac}.
When comparing the data with CCD quality lower spectra, there is considerable scatter in the CCD results, depending on how the spectra were analysed.

The \emph{Hitomi} results are unfortunately only for one cluster.
Making progress in understanding nucleosynthesis and the enrichment of clusters requires observations of more clusters with high spectral resolution.
In addition, we would like to be able to spatially resolve the cluster, to study how metals are transported through the intracluster medium, which is an indirect measurement of motions \cite{RebuscoDiff05}.
Using CCD-quality spectra maps of metallicity in clusters have been made (e.g. \cite{SandersCent16}), showing how they are transported in the cluster, sloshing with the surrounding hot gas, but having a patchy appearance.

\section{Beyond CIE plasma models}
\label{sect:beyondcie}
We have already discussed how resonant scattering can affect the spectra of clusters.
However, there are other phenomena which may be important in understanding these objects, which can only be addressed by high-resolution spectroscopy.

The non-baryonic process could contribute to the X-ray spectrum in clusters, leading to extra emission lines or modified spectra.
One interesting candidate is an emission line at $\sim 3.5$~keV found in stacked samples of clusters, or in some individual objects, including the Perseus cluster \cite{Bulbul14,Boyarsky14,Cappelluti18}, using low spectral resolution CCD data.
A possible origin for this line is that could originate from the radiative decay of a sterile neutrino.
This is particularly interesting, as such a particle is a candidate for dark matter \cite{Boyarsky19}.
However, such claims are controversial as some other analyses did not find similar lines (e.g. \cite{Tamura15}).
With lower spectral resolution CCD data there are a number of potential systematics which could  artificially create features which could be interpreted as lines.
Instrumental systematics uncertainties, such as the calibration of the effective area, have been suggested, although this is made less likely by the stacked objects lying at different redshifts.
Furthermore, the spectral model of the underlying cluster could introduce such features if there are lines missing or at the wrong strengths, leading to residuals at this energy.

To confirm the presence of the 3.5~keV line requires high spectral resolution data.
The line was not found at the expected strength in the \emph{Hitomi} spectrum of Perseus \cite{Hitomi352017}.
However, the flux of the line previously found in Perseus was much stronger than the average stacked signal (assuming an origin from dark matter), and so these data do not rule out a line at this level.
Further deep \emph{Hitomi}-quality spectra will undoubtedly help, although the line itself may be broad if its origin is from dark matter.

Another effect which may be generated by a different dark matter candidate, axion-like particles (ALPs), is an energy-dependent modulation of the spectra as photons and ALPs are interconverted \cite{Wouters13,Marsh17}.
The strength of this interaction varies with the magnetic field in the cluster as the photon travels through the cluster.
To look for the signal the spectrum of the cluster central AGN can be examined to look for variations from a power-law model.
CCD data analyses can be affected by various systematics (e.g. pileup, due to too many photons).
Deep \emph{Chandra} grating observations of Perseus were recently made \cite{Reynolds20}, placing very stringent limits on the strength of any possible ALP-photon coupling.
Future measurements, however, require excellent calibration of any used instrument to make further improvements.

An effect which may be present in the spectra of clusters is charge exchange (CX; Chapter 12).
This phenomenon occurs of a neutral atom interacts with a charged ion, where the ion then recombines into a highly excited state.
One region this may occur is in the filamentary systems seen in cool core clusters like Perseus \cite{Fabian11}.
In this model the hot gas penetrates the cooler filamentary gas through reconnection diffusion, producing soft X-ray emission.
Analysis of low resolution \emph{Chandra} spectra indicates CX could contribute to the X-ray flux in Perseus, but the RGS spectra indicate that is not the case in Centaurus \cite{WalkerFil15}.

CX has also been proposed to be the origin for the 3.5 keV emission line \cite{Gu15}, where it is due to a S~XVI line.
A search for CX O~VIII emission in stacked cluster RGS data found marginal evidence for this line \cite{Gu18}.

The assumption of a CIE equilibrium in the ICM of a galaxy cluster is also likely to be incorrect in some locations, for example in the outskirts of a galaxy cluster or in shocked regions \cite{FoxLoeb97}.
The level of non-equilibrium is important as it can bias important cluster observables, such as surface brightness, and physical quantities, such as hydrostatic mass and temperature (e.g. \cite{Camille15}).
Furthermore, it affects the spectrum in detail, allowing it to be detected from the anomalous line ratios in high-resolution spectra.
One which is important in clusters is the He-like O~VII triplet to H-like O~VIII doublet line ratio \cite{Wong11}.
Such a measurement could be difficult due to the surrounding complex warm-hot plasma and the requirements of future instrumentation.

\section{Future missions}
There are many ways that high-resolution spectroscopy can aid our understanding of galaxy clusters.
However, we are currently limited by instrumentation.
Our primary workhorse is the RGS grating detectors onboard \emph{XMM-Newton}.
These have produced amazing results, but to progress further, we need higher spectral resolution and more effective area.
In addition, the Fe-K lines cannot be examined using RGS, and these are particularly useful for velocity studies.

\emph{Hitomi} has shown the power of microcalorimetry in clusters.
These non-dispersive instruments are excellent for extended objects, because of the lack of spatial broadening of the spectra.
In addition, they have good spectral resolution over their entire range.
\emph{Hitomi} studied only one cluster, Perseus, but fortunately, the replacement mission \emph{XRISM} is due to be shortly launched (see Chapter 6).
This will have extremely similar capabilities to \emph{Hitomi}, allowing a much greater and more representative range of clusters and groups to be studied.
One particularly interesting target is the Coma cluster, where turbulence inside this merging cluster could be studied in detail \cite{ZuHone16}.

\begin{figure}
  \includegraphics[width=\textwidth]{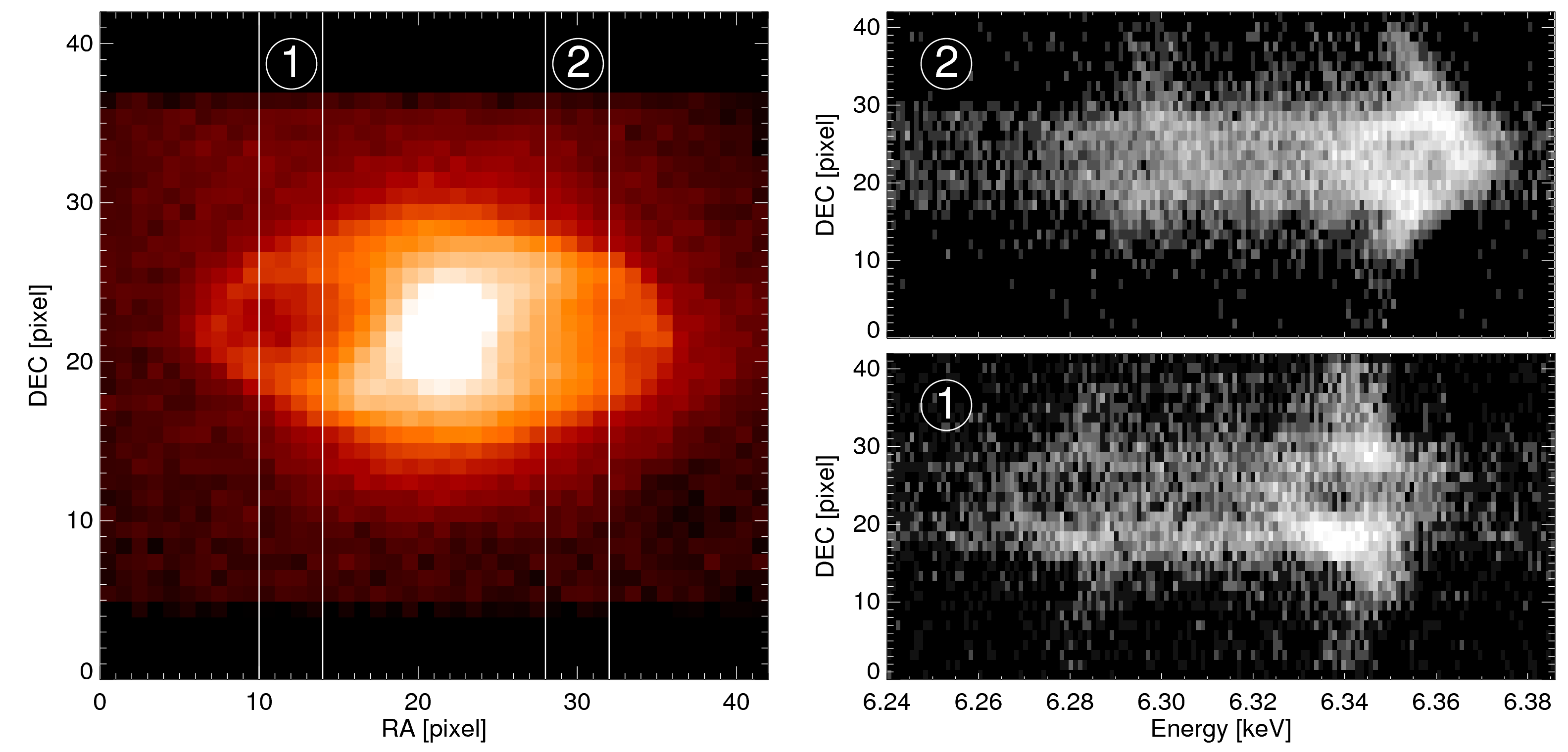}
  \includegraphics[width=\textwidth]{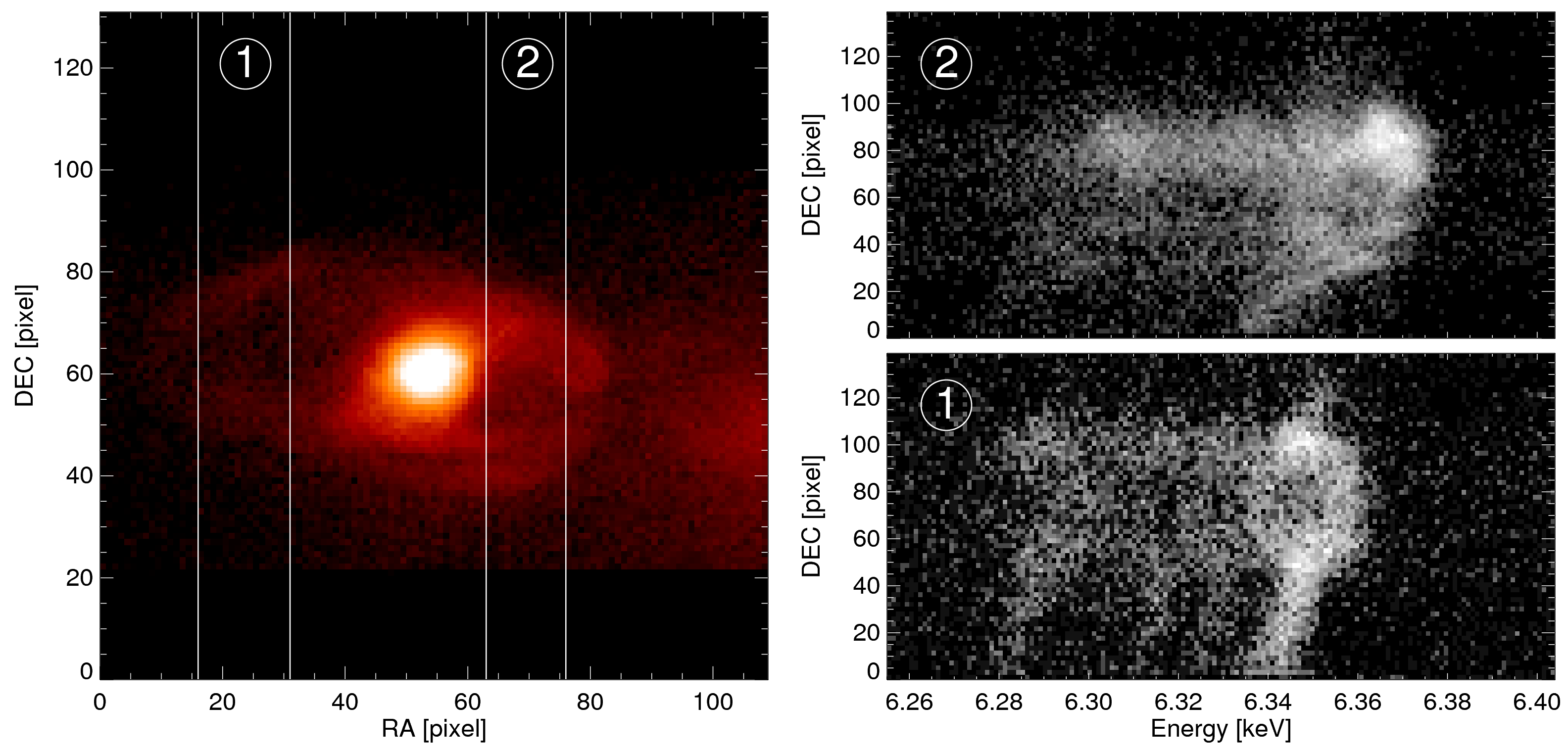}
  \caption{
    Simulations of AGN feedback of a Cygnus A-like cluster (top) and Hydra A (bottom) for the proposed \emph{Athena} X-IFU microcalorimeter array detector (taken from \cite{CrostonAthena13}) based on hydrodynamical simulations \cite{Heinz10}.
    The left panels show images on the detector and `virtual slit slices' labelled 1 and 2 across the cavities in the clusters.
    The right panels show the spectra around the 6.7 keV Fe~XXV K$\alpha$ along the virtual slits, for the 250~ks simulated observation.
    Using these spectra the velocity of the inflated bubble away and towards us can be inferred, along with its age and the jet power.
    In these simulations, the cavities have ages of 21 and 170~Myr, respectively.
  }
  \label{fig:athena_feedback}
\end{figure}

Although \emph{XRISM} will make great advances, further future progress will also require a larger effective area and better spatial resolution.
Spatial resolution is required to study in detail the different regions inside a cluster.
In the core, in particular, we would like to study in detail the individual regions surrounding the central AGN to understand feedback.
One mission that is hoped to provide these capabilities is the proposed \emph{Athena} observatory \cite{NandraAthena13}.
The currently-designed X-IFU detector on \emph{Athena} would provide $\sim 5$~arcsec spatial resolution combined with a large effective area.
Fig.~\ref{fig:athena_feedback} shows simulations of AGN feedback in two galaxy clusters as observed by an X-IFU-like instrument.
It shows that the shape of the Fe-K lines can be examined in detail in different locations around the cluster, allowing the velocity of the material surrounding the inflated bubble to be directly measured.
In addition, \emph{Athena} would be able to measure turbulence and motions generated in the surrounding region.
Such measurements would be extremely powerful in understanding the feedback process.

\begin{figure}
  \includegraphics[width=\textwidth]{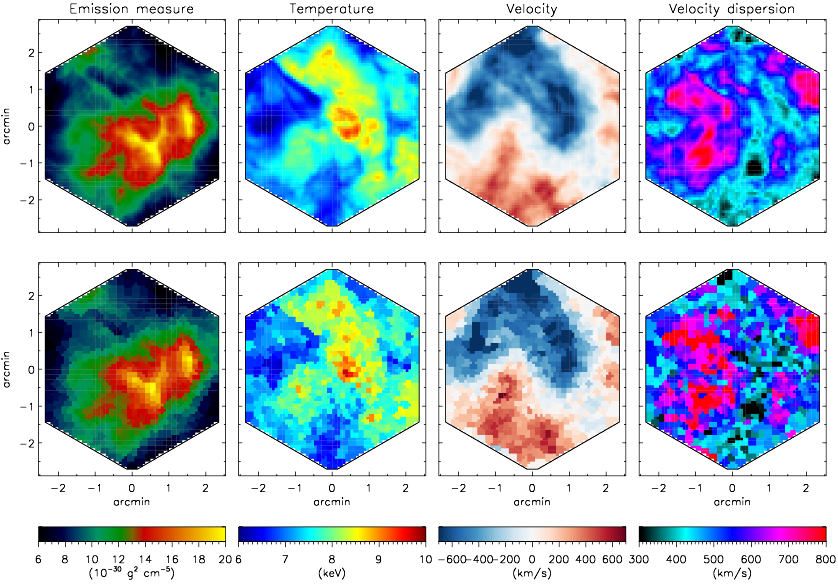}
  \caption{
    Maps of input properties (top) and recovered values (bottom) for a simulated X-IFU observation (taken from \cite{Roncarelli18}).
  }
  \label{fig:athena_maps}
\end{figure}

\emph{Athena} will also allow detailed maps of the motions in clusters to be made, including the velocity and velocity dispersion.
Fig.~\ref{fig:athena_maps} shows maps of a Coma-like galaxy cluster, where input maps are simulated and recovered \cite{Roncarelli18}.
In these simulated data, the 2D power spectrum was obtained, which is an important measurement to with which turbulence can be studied.
The higher signal-to-noise ratio \emph{Athena} would produce also has a big advantage in distinguishing studying multiple velocity components along the line as sight, as occurs in a major merger \cite{Biffi22}.

Metallicity maps made with CCD detectors have shown that metals are not smoothly distributed through the cluster (e.g. \cite{SandersCent16}), but they sloshed around and appeared patchy.
Detailed studies with spatially resolved high-resolution spectroscopy would be invaluable due to the complex temperature structure of these objects.
\emph{Athena} X-IFU should have the capability of mapping temperature, metallicity and density out to a large radius in a galaxy cluster, studying chemical evolution in clusters and metal transport through the ICM \cite{Cucchetti18}.

\bibliographystyle{myspringer}
\bibliography{refs.bib}

\end{document}